\documentclass[10pt,journal,epsfig,twosides]{IEEEtran}
\usepackage{amsmath}
\usepackage{graphicx}
\usepackage{amssymb}
\usepackage{amsfonts}
\usepackage{cite}
\usepackage{bm,comment}

\usepackage{subeqnarray}
\usepackage{subfigure}
\usepackage{multirow} 
\usepackage{multicol}
\usepackage{slashbox}
\usepackage{stfloats}
\usepackage{float}
\usepackage{color}
\usepackage{epstopdf}
\makeatletter
\newif\if@restonecol
\makeatother

\usepackage[ruled,vlined]{algorithm2e}
\usepackage{algpseudocode}
\usepackage{amsmath}

\newtheorem{proposition}{Proposition}
\newtheorem{remark}{Remark}

\newtheorem{lemma}{Lemma}
\newtheorem{corollary}{Corollary}

\newcommand{\tabincell}[2]{\begin{tabular}{@{}#1@{}}#2\end{tabular}}

\begin{document}
\IEEEoverridecommandlockouts
\title{Terahertz Multi-User Massive MIMO with Intelligent Reflecting Surface: Beam Training and Hybrid Beamforming }
\author{Boyu Ning, \IEEEmembership{Student Member, IEEE}, Zhi Chen, \IEEEmembership{Senior Member, IEEE}, \\Wenrong Chen, Yiming Du, and Jun Fang, \IEEEmembership{Senior Member, IEEE}
\thanks{Copyright (c) 2015 IEEE. Personal use of this material is permitted. However, permission to use this material for any other purposes must be obtained from the IEEE by sending a request to pubs-permissions@ieee.org.}
\thanks{Part of this paper been presented in the IEEE  International Conference on Communications (ICC) 2020\cite{ori}.}
\thanks{This work was supported in part by the National Key R$\&$D Program of China under Grant 2018YFB1801500.}
\thanks{B. Ning, Z. Chen, W. Chen, Y. Du, and J. Fang are with National Key Laboratory of Science and Technology on Communications, University of Electronic Science and Technology of China (UESTC), Chengdu 611731, China (e-mails: boydning@outlook.com; chenzhi@uestc.edu.cn; wenrongchen@std.uestc.edu.cn; duyiming01009@163.com; junfang@
uestc.edu.cn).}}
\maketitle

\begin{abstract}
Terahertz (THz) communications open a new frontier for the wireless network thanks to their dramatically wider available bandwidth compared to the current micro-wave and forthcoming millimeter-wave communications. However, due to the short length of THz waves, they also suffer from severe path attenuation and poor diffraction. To compensate for the THz-induced propagation loss, this paper proposes to combine two promising techniques, viz., massive multiple input multiple output (MIMO) and intelligent reflecting surface (IRS), in THz multi-user communications, considering their significant beamforming and aperture gains. Nonetheless, channel estimation and low-cost beamforming turn out to be two main obstacles to realizing this combination, due to the passivity of IRS for sending/receiving pilot signals and the large-scale use of expensive RF chains in massive MIMO. In view of these limitations, this paper first develops a cooperative beam training scheme to facilitate the channel estimation with IRS. In particular, we design two different hierarchical codebooks for the proposed training procedure, which are able to balance between the robustness against noise and searching complexity. Based on the training results, we further propose two cost-efficient hybrid beamforming (HB) designs for both single-user and multi-user scenarios, respectively. Simulation results demonstrate that the proposed joint beam training and HB scheme is able to achieve close performance to the optimal fully digital beamforming  which is implemented even under perfect channel state information (CSI).
\end{abstract}

\IEEEpeerreviewmaketitle

\section{Introduction}
Data rate demands in wireless communications have soared rapidly in the last few decades. Due to this dramatically increasing demand, the available bandwidth in the sub-6 gigahertz (GHz) and millimeter-wave (mmWave) bands will become inadequate to support various bandwidth-hungry applications in the future, such as virtual reality, augmented reality, and so on. Recently, terahertz (THz) communication has emerged as an appealing technology to cope with the above bandwidth ``crisis'', by shifting the current frequency bands to the under-exploited and wide frequency band over 0.1-10 THz \cite{IF,Zchen}. Despite the orders-of-magnitude increase in the bandwidth, THz communications inevitably possess two major shortcomings owing to the short length of THz waves, i.e., \emph{severe signal attenuation} and \emph{poor diffraction}\cite{spr}, which render the THz signals very susceptible to obstacles, thus severely limiting their coverage \cite{hancong}.

To break this short-range bottleneck and realize uninterrupted wireless connectivity, this paper proposes to combine two promising techniques, namely, massive multiple-input multiple-output (MIMO)\cite{larsson} and intelligent reflecting surface (IRS)\cite{MDI} in THz communication, considering their appealing potentials for signal enhancement. Specifically, massive MIMO can compensate for the severe signal attenuation by generating narrow beams with strong beamforming gains, whereas IRS can provide additional aperture gains via controllable signal reflection.  By mounting the IRS at suitable positions for signal enhancement, the THz signal blockage can be significantly reduced.  

\subsection{Prior Works}\label{PW}
Thanks to the small wavelength in the THz frequency band, more antenna elements can be packed in an antenna array with half-wavelength spacing, thus facilitating the use of massive MIMO in THz communication for expanding the communication distance and enhancing achievable data rates\cite{chan,hsa}. Nonetheless, the increased transmit/receive dimensions also complicate the signal processing and scale up the hardware costs, as each antenna requires a dedicated radio frequency (RF) chain and an analog-to-digital converter (ADC) in fully-digital beamforming architecture\cite{sun}. To reduce the hardware cost,  hybrid beamforming (HB) that implements part of MIMO beamforming in the analog domain only using phase shifts (PSs) has been considered as a viable alternative\cite{aal}. In particular, since the THz channel can be modeled as the accumulation of multiple propagation paths with different angles of departure/arrival (AoD/AoA) and path loss\cite{hancong}, one common HB design is by setting the analog precoder/combiner as the array response vector (ARV) corresponding to the directions of the propagation paths\cite{eud4,gao} due to the property of beam orthogonality in massive MIMO\cite{5}, while leaving the digital precoder/combiner (yet with reduced computational dimension compared to fully digital beamforming) for optimization only. By this means, the overall complexity of beamforming design can be greatly decreased.

On the other hand, IRS is a novel low-complexity hardware technology which have been utilized to realize transceiver or relay architectures\cite{ZL,w1,w2}. Most of the existing works on IRS aim to optimize its PSs for enhancing communication performance. For example, the authors in \cite{huangchi,huangchi2,huangchi3} aimed to maximize the network spectral efficiency and energy efficiency, under the assumption that zero-forcing beamforming is applied at the base station (BS). Furthermore, the authors in \cite{qinte,qinte2,qinte3,qinte4} considered joint BS beamforming and IRS-PSs optimization for maximizing the total received signal power at the destination.  Similar design problems have also been investigated for improving different metrics, such as the minimum signal-to-noise-plus-interference ratio (SINR)\cite{ism},  multi-user weighted sum-rate\cite{ghy}, secrecy rate in physical layer security\cite{cuim}, and so on. Nonetheless, the above works focused merely on the single/multiple-input single-output (SISO/MISO) system. The investigation on the more challenging scenario of IRS-assisted MIMO is still limited. In \cite{nby} and \cite{zhangs}, two suboptimal BS beamforming and IRS-PSs designs were proposed by applying the sum-path-gain maximization (SPGM) criterion and alternating optimization algorithm, respectively, assuming a point-to-point MIMO system. In \cite{cpan}, authors jointly optimized the BS beamforming and IRS-PSs in a multi-cell communication network via the block coordinate descent algorithm. 

It is worth noting that all of the aforementioned works \cite{huangchi,huangchi2,huangchi3,qinte,qinte2,qinte3,qinte4,ism,ghy,cuim,nby,zhangs, cpan} assumed perfect channel state information (CSI) knowledge on all links involved. However, the CSI acquisition is a non-trivial issue in IRS-assisted wireless communications since IRS cannot actively transmit/receive signals as BS or user due to the lack of RF chains. Although several works have looked into new channel estimation schemes for the IRS-assisted MIMO systems recently\cite{est1,est2,est3}, it is practically inefficient to combine the channel estimation schemes in \cite{est1,est2,est3} with the beamforming designs in \cite{nby,zhangs,cpan} in massive MIMO systems, as their proposed schemes are hardly scalable with the number of transmit/receive antennas, which could result in extremely high implementation complexity. In this regard, it is of great importance to investigate a new, effective and low-complexity channel estimation and beamforming strategy for THz massive MIMO systems. It is worth noting that there have been some initial works focusing on minimizing the training overheads induced by IRS-assisted massive MIMO systems\cite{z1,z2}. However, due to the severe path-loss in THz channels, the receive signals may be too weak to be detected by these strategies since a crucial beam alignment issue is not considered in \cite{z1} and \cite{z2}. 

 \subsection{Main Contributions}\label{CO}
 Motivated by the above, in this paper, we propose a low-complexity \emph{beam training strategy} for multi-user IRS-assisted massive MIMO systems with an HB architecture in THz communications, so as to facilitate both efficient channel estimation and beamforming design. It is worth noting that in the conventional beam training scheme without IRS,  the channel information can be estimated by exhaustively searching over all available AoD/AoA beam combinations at BS/users. However, this method is ineffective in our considered system with IRS, as IRS cannot generate or receive beams by itself. Our main contributions are summarized as below.\begin{itemize}
 \item We first present an exhaustive beam training method to acquire the optimal AoDs/AoAs combination for the direct line-of-sight (LoS) links and reflecting links, respectively. In particular, we propose a new beam training criterion that applies the beams with the same coverage-edge gain instead of the conventional uniformly distributed beams (i.e., the beams with the same beamwidth), which is shown able to reduce the misalignment probability.
 \item Furthermore, owing to the overly high searching complexity of the exhaustive method, we propose a low-complexity cooperative beam training method that combines partial search at IRS-PSs and hierarchical search at BS and users. Specifically, we first show that with the proposed scheme, there is no need to search the exact path AoD/AoA at IRS itself but only need to search the angle differences in the sine space with fewer codewords. To implement the beam search, we propose a ternary-tree hierarchical beam search at BS and users, which is proved to be more efficient compared to the binary-tree search used in the conventional beam training\cite{track1,track2,track3,track4}.
\item To realize the ternary-tree search, we design two novel training codebooks, viz.,  tree dictionary (TD) codebook and  PS deactivation (PSD) codebook. These two codebooks share the same bottom-stage narrow beams but differ in upper-stage wide beams, so as to cater to different application scenarios.  In particular, the TD codebook has higher anti-noise capability but requires all RF chains in the realization of each wide beam, whereas the PSD codebook is more sensitive to noise but only requires one RF chain per wide beam, which thus shortens the search time by enabling simultaneous test of multiple beams.
\item Based on the training results, the IRS-PS and the analog precoder/combiner can be determined straightforwardly as the optimal IRS codeword and the ARVs with corresponding AoD/AoA\cite{eud4,gao}, respectively. To determine the digital precoder/combiner, we first consider the single-user scenario, for which the analog precoder/combiners are naturally close to the singular vectors of the effective channel between the BS and the user with the IRSs. As such, the digital precoder/combiner design is approximately equivalent to the direct power allocation (DPA) over the transmit beams via water-filling. However, for the multi-user scenario, the DPA method may result in inter-user interference and thus impair the sum-rate performance. As such, we apply the block diagonalization (BD) on the effective channels between all users' digital combiner and the BS's digital precoder, so as to maximize the sum rate while eliminating the inter-user interference.
\item Simulation results demonstrate that the proposed ternary-tree beam training schemes by the two codebooks can both realize $100\%$ detection probability in the high signal-to-noise ratio (SNR) regime. Moreover, the proposed HB designs under the estimated CSI are able to achieve close performance to the optimal fully digital beamforming which is implemented even under perfect CSI.
\end{itemize}
 
\subsection{Organization and Notations}
The remainder of the paper is organized as follows. The system model is introduced in Section II.
The exhaustive beam training method and the proposed low-complexity beam training method are presented in Section III and Section IV, respectively.  The proposed IRS-PSs and HB designs are presented in Section V. Numerical results are provided in Section VI, and conclusions are finally drawn in Section VII.

We use small normal face for scalars, small bold face for vectors, capital bold face for matrices, $\mathbb{C}$ for complex numbers, and ${\mathbb{N}}+$ for positive integers. ${\mathop{\mathbb{E}}} {\rm{\{ }} \cdot \}$, $\det ( \cdot )$, $|\cdot|$, and ${\left\| \cdot \right\|_p}$ represent expectation, determinant, modulus, and $p$-norm, respectively. The superscript ${{\rm{\{ }} \cdot \}^T}$, ${{\rm{\{ }} \cdot \}^{\dag}}$, and ${{\rm{\{ }} \cdot \}^H}$ denote the transpose, conjugate, and Hermitian transpose, respectively. ${\bf{a}}(n)$ and ${\bf{A}}(:,n)$ represent the $n$th element and the $n$th column of  ${\bf{a}}$ and ${\bf{A}}$, respectively.  ${\rm{diag}}( \cdot )$ denotes a diagonal matrix whose diagonal elements are given by its argument. $\otimes$ and  $\odot$ are Kronecker product and Hadamard product operators, respectively. $\left\lfloor  \cdot  \right\rfloor$ is the floor integer function and $\left\lceil  \cdot  \right\rceil$ is the ceiling integer function. ${\rm{span}}( \cdot )$ denotes the subspace spanned by the collection of vectors and $\bot$ denotes the orthogonal complement of a subspace. $\mathcal{CN}(\mu,\sigma^2)$ means circularly symmetric complex Gaussian (CSCG) distribution with mean of $\mu$ and variance of $\sigma^2$. ${{\bf{I}}_N}$ denotes the $N \times N$ identity matrix. 

\section{System Model}
\begin{figure}[t]
\centering
\includegraphics[width=3.5in]{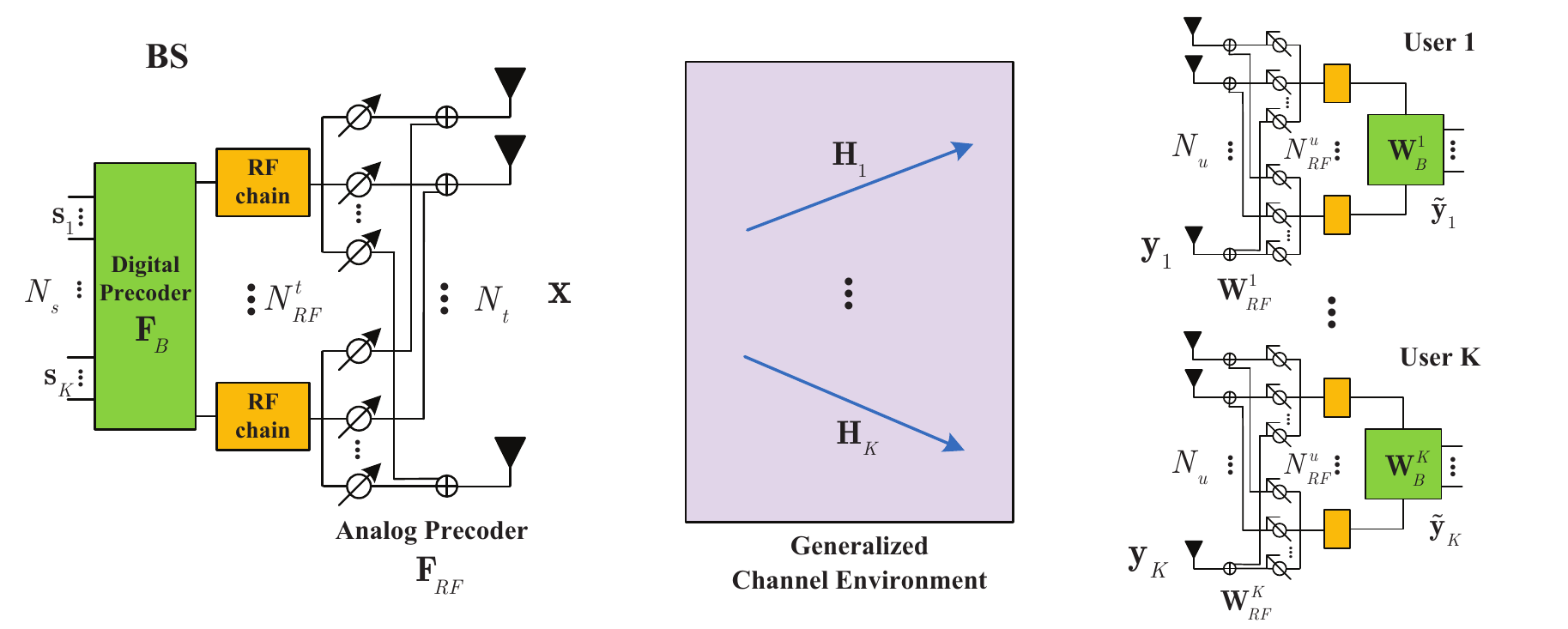}
\caption{A multi-user massive MIMO system with the HB architecture at the BS and the user sides.}\label{figmimo}
\vspace{-10pt}
\end{figure}
\subsection{Massive MIMO Architecture with HB}
We consider a narrowband downlink multi-user massive MIMO system as depicted in Fig. \ref{figmimo}, where a BS with $N_t$ antennas intends to send $N_s$ data streams to $K$ users, each with $N_u$ antennas and $D_k$ required data symbols and ${N_s} = \sum\nolimits_{k = 1}^K {{D_k}}$. The results in this paper can be also extensible to the uplink. A transmit vector ${\bf{x}} \in {\mathbb{C}^{{N_t}}}$ is generated at the BS via the following HB process. First, a data stream vector ${\bf{s}} \in {C^{{N_s}}}$ is precoded by a baseband digital precoder ${{\bf{F}}_{B}} \in {\mathbb{C}^{{{N_{RF}^t} \times {N_s}}}}$, and then up-converted to the RF domain by passing through $N_{RF}^t$ RF chains before being precoded by an analog precoder ${{\bf{F}}_{RF}} \in {\mathbb{C}^{{N_t} \times {N_{RF}^t}}}$. Each column of ${{\bf{F}}_{RF}} $ can be selected from a set of ARVs implemented by PSs, i.e.,
\begin{equation}\label{pran}
{{\bf{F}}_{RF}}(:,i) \in \mathcal{F}_{\rm{RF}}=\{ {{\bf{a}}_{{N_t}}}({\varphi _1}),...,{{\bf{a}}_{{N_t}}}({\varphi _{N}})\},\; \forall i,
\end{equation}
where $\{{{\bf{a}}_{N_t}}(\varphi_n)\}_{n=1}^N$ represent $N$ predefined ARVs. For ease of exposition, we consider a uniform linear array configuration and the normalized ARV with $N_a \in \{N_t,N_u\}$ antenna elements is given by
\begin{equation}\label{ula}
{{\bf{a}}_{N_a}}(\phi ) = \frac{1}{{\sqrt {N_a} }}{[1,{e^{jkd_a\sin (\phi )}},...,{e^{jkd_a(N_a - 1)\sin (\phi )}}]^T},
\end{equation}
where $k = 2\pi /\lambda $, $\lambda$ is the wavelength, $d_a$ is the antenna spacing, and $\varphi$ denotes the AoD or AoA. Considering a normalized power constraint ${\| {{{\bf{F}}_{RF}}{{\bf{F}}_{B}}} \|_2^2} = 1$, the transmitted signals by HB can be written as
\begin{equation}\label{1}
{\bf{x}} = \sqrt P {{\bf{F}}_{RF}}{{\bf{F}}_{B}}{\bf{s}} = \sqrt P \sum\limits_{i = 1}^K {{{\bf{F}}_{RF}}} {{\bf{F}}_{B}^i}{{\bf{s}}_i},
\end{equation}
where $P$ is the total transmit power of the BS, ${{\bf{F}}_{B}^i}\in {\mathbb{C}^{{N_{RF}^t}\times{D_i}}}$ and ${{\bf{s}}_i}\in {\mathbb{C}^{{D_i}}}$ are the subprecoders and the data stream vector for user$i$, i.e., ${{\bf{F}}_B} = [{\bf{F}}_B^1,{\bf{F}}_B^2,...,{\bf{F}}_B^K]$ and ${\bf{s}} = {[{\bf{s}}_1^T,{\bf{s}}_2^T,...,{\bf{s}}_K^T]^T}$. Here, we assume $E[{\bf{s}}{{\bf{s}}^H}] = {{\bf{I}}_{{N_s}}}$. For user$k$, its received signal can be expressed as
\begin{equation}
{{\bf{y}}_k} = \sqrt P {{\bf{H}}_k}\Big( {\underbrace {{{\bf{F}}_{RF}}{{\bf{F}}_{B}^k}{{\bf{s}}_k}}_{\text{desired\;signal  \;}} + \underbrace {\sum\limits_{i \ne k}^K {{{\bf{F}}_{RF}}} {{\bf{F}}_{B}^i}{{\bf{s}}_i}}_{{\text{interference\;signal \;}}}} \Big) + {\bf{n}},
\end{equation}
where ${{\bf{H}}_k} \in {{\mathbb{C}}^{{N_u} \times {N_t}}}$ is the channel matrix from the BS to user$k$, ${\bf{n}} \in {\mathbb{C}^{{N_u}}}$ is additive white Gaussian noise following $\mathcal{CN}(\mathbf{0},{\sigma _n^2}{\bf{I}}_{N_u})$ with $\sigma_n^2$ denoting the noise power for each user antenna element. At user$k$, the received signal ${\bf{y}}_k \in {\mathbb{C}}^{{N_u}}$ is first processed by an analog combiner ${\bf{W}}_{RF}^k \in {\mathbb{C}}^{{N_u} \times {N_{RF}^{u,k}}}$ selected from $N$ predefined antenna ARVs, i.e.,
\begin{equation}\label{coman}
{\bf{W}}_{RF}^k(:,i) \in \mathcal{W}_{\rm{RF}}=\{ {{\bf{a}}_{{N_u}}}({\varphi _1}),...,{{\bf{a}}_{{N_u}}}({\varphi _{N}})\},\;\forall i,
\end{equation}
and then down-converted to the digital domain via $N_{RF}^{u,k}$ RF chains and combined by a baseband digital combiner ${\bf{W}}_{B}^k \in {\mathbb{C}}^{{N_{RF}^{u,k}} \times {D_k}}$, which results in
\begin{equation}\label{shoufa}
{\widetilde {\bf{y}}_k} = \sqrt P {{\bf{W}}_k^H}{{\bf{H}}_k}{{\bf{x}}_{ds}} + \underbrace {\sqrt P {{\bf{W}}_k^H}{{\bf{H}}_k}{{\bf{x}}_{is}} + {{\bf{W}}_k^H}{\bf{n}}}_{{\text{effective}}\;{\text{noise}}},
\end{equation}
where ${{\bf{x}}_{ds}} = {{\bf{F}}_{RF}}{{\bf{F}}_{B}^i}{{\bf{s}}_i}$, ${{\bf{x}}_{is}} = \sum\nolimits_{i \ne k}^K {{{\bf{F}}_{RF}}} {{\bf{F}}_{B}^i}{{\bf{s}}_i}$, and ${{\bf{W}}_k} = {\bf{W}}_{RF}^k{\bf{W}}_{B}^k$. 
\subsection{THz Channel Model with IRS} 
\begin{figure}[t]
\center
\includegraphics[width=2.8in]{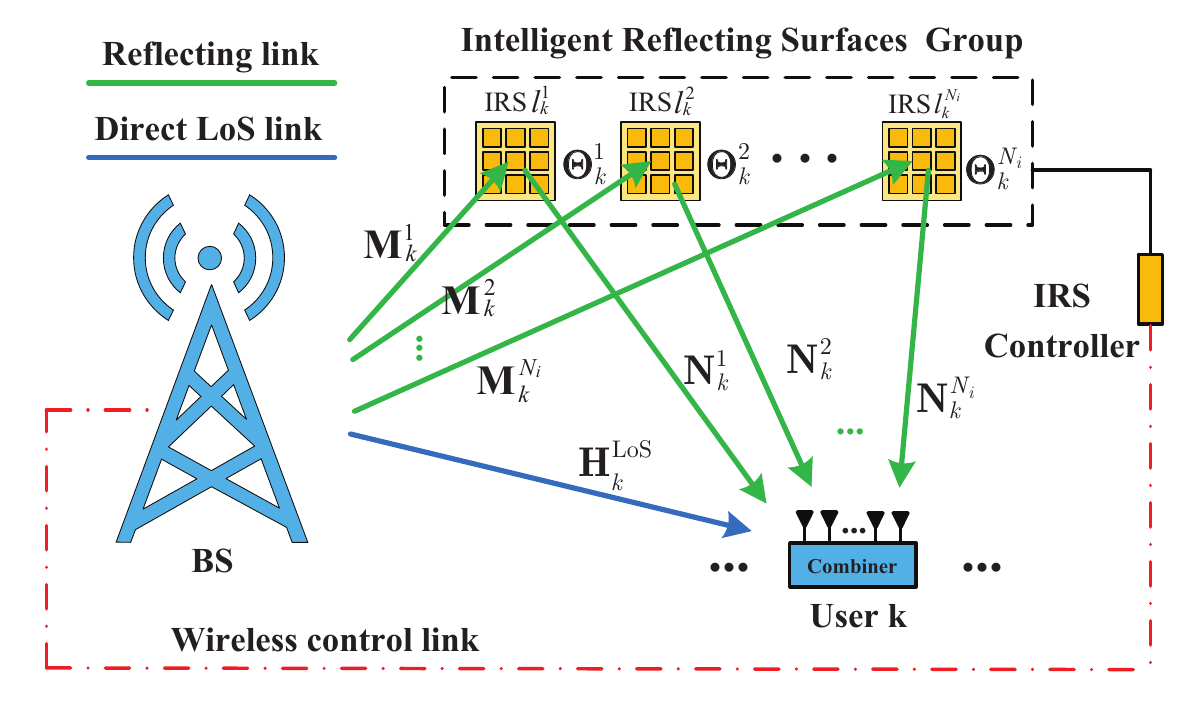}
\caption{Illustration of the channel between the BS and user$k$, i.e., ${\bf{H}}_k$.}\label{figirs}
\vspace{-12pt}
\end{figure} 
We consider that each user is served by $N_i$ IRSs which are installed on a surrounding wall to enhance its communication quality with the BS, as shown in Fig. \ref{figirs}, where $l_k^1,l_k^2,...,l_k^{N_i}$ denote the indices of the IRSs serving user$k$. Note that in THz communication, the power of scattering component is generally much lower (more than 20 dB) than that of LoS component\cite{model}. Thus, we ignore the scattering component and the channel between the BS and user$k$ should be dominated by the LoS component and can be expressed as
\begin{equation}\label{chan}
{{\bf{H}}_k}(f,{{\bf{d}}_k})\!=\! G_tG_r\Big[{\bf{H}}_k^{\rm{LoS}}(f,{d_{k,0}}) \!+\! \sum\limits_{m = 1}^{N_i} {{\bf{H}}_{k,m}^{\rm{Ref}}(f,d_{k,m}^M,d_{k,m}^N)}\Big],
\end{equation}
where $G_t$ and $G_r$ are the transmit and receive antenna gains, respectively, ${\bf{H}}_k^{\rm{LoS}}$ is the BS-user$k$ direct LoS channel, $\{{\bf{H}}_{k,m}^{\rm{Ref}}\} _{m = 1}^{N_i}$ are $N_i$ BS-IRS-uesr$k$ cascaded channels, $f$ is the carrier frequency, and ${{\bf{d}}_k}=[d_{k,0},d_{k,1}^M,d_{k,1}^N,..., d_{k,N_i}^M,d_{k,N_i}^N]$ is the path distance vector, with $d_{k,m}^M$ (resp. $d_{k,m}^N$) denoting the distance between the BS and IRS$m$ (resp. IRS$m$ and user$k$). Each IRS can dynamically adjust its $N_r$ reflecting elements  to shift the phase of the incoming signals in a passive manner (i.e., without energy amplification). As such, the received signal at IRS$m$ can be linearly transformed by a diagonal PS matrix ${\bf{\Theta }}_k^m \triangleq {\rm{diag}}(\beta {e^{j{\theta _1^{k,m}}}},\beta {e^{j{\theta _2^{k,m}}}}, \cdots ,\beta {e^{j{\theta _{{N_r}}^{k,m}}}}),\;m=1,2,...,N_i,$ and reflected to user$k$, where $j = \sqrt { - 1}$, $\{ {\theta _i^{k,m}}\} _{i = 1}^{{N_r}} \in [0,2\pi )$, and $\beta \in [0,1]$ are imaginary unit, PSs of IRS$m$ which serves user$k$, and amplitude reflection coefficient, respectively. Thus, the $N_i$ reflecting links between the BS and user$k$ due to the $N_i$ IRSs in (\ref{chan}) can be expressed as
\begin{equation}\label{chan2}
\sum\limits_{m = 1}^{{N_i}} {{{\bf{H}}_{k,m}^{{\rm{Ref}}}}(f,d_{k,m}^M,d_{k,m}^N)}  = \sum\limits_{m= 1}^{{N_i}} {{\eta }{{\bf{N}}_k^m}{{\bf{\Theta }}_k^m}} {{\bf{M}}_k^m},
\end{equation}
where $\eta$ is the path-loss compensation factor (to be specified in (\ref{comp})), ${{\bf{M}}_k^m}$ is the BS-IRS$l_k^m$ LoS channel, and ${{\bf{N}}_k^m}$ is the IRS$l_k^m$-user$k$ LoS channel. Without out loss of generality, we assume all the IRSs are linear arrays for ease of exposition. As a result, the LoS channels in (\ref{chan}) and (\ref{chan2}) can be more explicitly written as a combination of ARVs, i.e.,
\begin{align}
{\bf{H}}_k^{\rm{LoS}}(f,{d_{k,0}}) &= a(f,{d_{k,0}}){\bf{a}}_{{N_u}}^U\left( {\varphi _{U,H}^k} \right){\bf{a}}_{{N_t}}^B{\left( {\varphi _{B,H}^k} \right)^H},\label{sepchan}\\
{{\bf{M}}_k^m}(f,d_{k,m}^M) &= a(f,d_{k,m}^M){\bf{a}}_{{N_r}}^R\left( {\varphi _{R,M}^{k,m}} \right){\bf{a}}_{{N_t}}^B{\left( {\varphi _{B,M}^{k,m}} \right)^H},\notag\\
{\bf{N}}_k^m(f,d_{k,m}^N) &= a(f,d_{k,m}^N){\bf{a}}_{{N_u}}^U\left( {\varphi _{U,N}^{k,m}} \right){\bf{a}}_{{N_r}}^B{\left( {\varphi _{R,N}^{k,m}} \right)^H},\notag
\end{align}
where ${\varphi}$ is the AoA/AoD angle\footnote{The subscripts U, R, and B represent the user, IRS, and BS, respectively, and the subscripts H, M, and N represent the channel ${\bf{H}}$, ${\bf{M}}$, and ${\bf{N}}$, respectively. The superscripts $k$ and $k,m$ represents the BS-user$k$ link and the BS-IRS$l_m^k$-user$k$ link, respectively.}. $a(f,d)$ represents the path loss consisting of the free-spread loss and the molecular absorption loss in THz communication and satisfies\cite{hancong}
\begin{equation}\label{loss}
a(f,d) = {\frac{c}{{4\pi fd}}}{e^{-\frac{1}{2}\tau (f)d}},
\end{equation}
where $c$ is the speed of light and $\tau (f)$ is the medium absorption factor. According to \cite{gain}, the maximum energy of the reflected signal over the BS-IRS$l_k^m$-user$k$ link is given by
\begin{equation}\label{tm}
{P_m} = \frac{{PG{{({\varphi _{R,M}^{k,m}})}^2}N_r^2{c^2}d_{\rm{IRS}}^2{\beta ^2}}}{{64{\pi ^3}{{(d_{k,m}^Md_{k,m}^N)}^2}{f^2}}}{e^{ - \tau (f)(d_{k,m}^M + d_{k,m}^N)}},
\end{equation}
where $G$ is the IRS element gain and $d_{\rm{IRS}}$ is the side length of a square IRS element. $G(\varphi)^2$ describes how much power is transmitted or received in direction $\varphi$ to that of an isotropic antenna, i.e., \\
\begin{equation}
G(\varphi) = \sqrt {F(\varphi)\frac{{2\pi }}{{ {\int\limits_{\varphi  = -\pi}^{\pi}  {F(\varphi )d\varphi} } }}},
\end{equation}
and $F(\varphi)$ is the function of normalized power radiation pattern. As for our cascaded channel model in (\ref{chan2}), the reflected signal over the BS-IRS$l_k^m$-user$k$ link is written as $\eta{\bf{N}}_k^m{\bf{\Theta }}_k^m{\bf{M}}_k^m{\bf{x}}$ and the maximum energy is given by
\begin{equation}\label{om}
\begin{split}
{P_m} = P{[{N_r}\eta \beta a(f,d_{k,m}^M)a(f,d_{k,m}^N)]^2}.
\end{split}
\end{equation}
As such, by equalizing (\ref{tm}) and (\ref{om}), the path-loss compensation factor is given by
\begin{equation}\label{comp}
\eta  = \frac{{2\sqrt \pi  G({\varphi _{R,M}^{k,m}}) d_{{\rm{IRS}}}^2}}{{{\lambda ^2}}}.
\end{equation}
Practically, the side length of the IRS element $d_{{\rm{IRS}}}$ is within the range $[\frac{\lambda}{10},\frac{\lambda}{2}]$. Typically, for a half-space isotropic pattern, $F(\varphi)$ can be simply expressed as
\begin{equation}
F\left( {\varphi} \right) = \left\{ {\begin{split}
1,&{}&{\varphi \in \left[ {-\frac{\pi }{2},\frac{\pi }{2}} \right]}\\
0,&{\quad}&{\rm{otherwise}\quad}
\end{split}} \right..
\end{equation}
In this case, we have $G=\sqrt{2}$ and hence $\eta  \in [\frac{{\sqrt {2\pi } }}{{50}},\frac{{\sqrt {2\pi } }}{2}]$. 
\section{Exhaustive Beam Training Method}
 \begin{figure}[t]
\center
\includegraphics[width=2.5in]{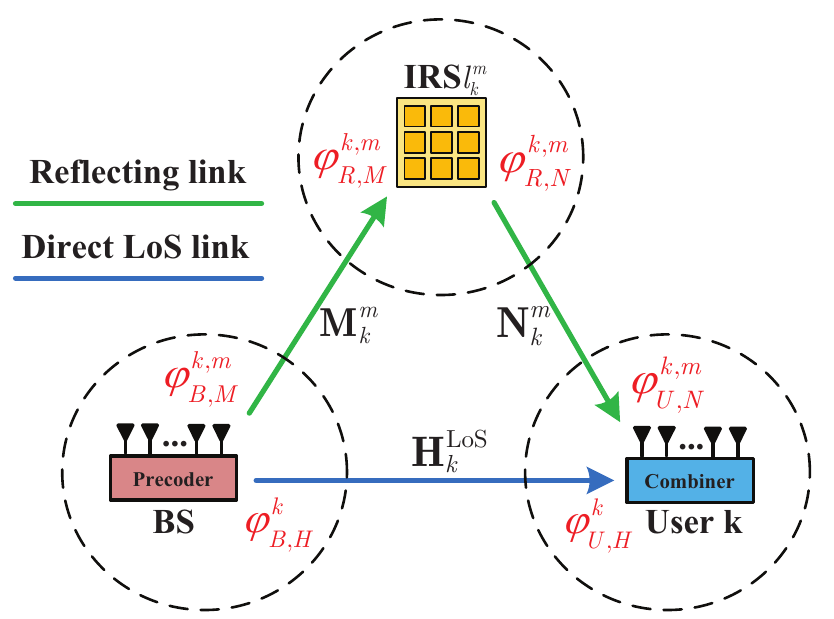}
\caption{A basic three-node communication model of the IRS-assisted systems.}\label{fig3n}
\vspace{-12pt}
\end{figure}
For the geometric THz channel in (\ref{sepchan}),  the main aim of channel estimation is to secure the path parameters, i.e.,  AoAs,  AoDs, and path loss. As shown in Fig. \ref{fig3n},  a basic three-node communication model consists of six path angles  $\varphi _{B,H}^k,\varphi _{U,H}^k,\varphi _{B,M}^{k,m},\varphi _{R,M}^{k,m}, \varphi _{R,N}^{k,m},$ and $\varphi _{U,N}^{k,m}$ to be estimated.  In the rest of this section, we develop an exhuastive beam training method to find the path angles for both the direct LoS link and the reflecting link.

\subsection{Beam Training for Direct LoS Link}\label{co}
To estimate the angles $\varphi _{B,H}^k$ and $\varphi _{U,H}^k$ for the BS-user$k$ path (see Fig. \ref{fig3n},), we can define $N$ narrow beams $\{{\bf{a}}_{N_a}(\varphi_n)\}_{n=1}^N$ in (\ref{pran}) and (\ref{coman}) with $N\ge N_a$ that cover the whole space and search an optimal beam pair by enumerating $N \times N$ narrow-beam combinations\cite{track2}. By this means, $\varphi _{B,H}^k$ and $\varphi _{U,H}^k$  can be estimated accordingly based on the directions of the optimal beam pair. Different from \cite{track2} which assumes that the $N$ narrow beams are uniformly distributed within $[0,2\pi]$, we first show that it suffices to consider these beams only within $[ - \frac{\pi }{2},\frac{\pi }{2}]$ due to the following lemma.
\begin{lemma}
Beams within $[ - \frac{\pi }{2},\frac{\pi }{2}]$ and beams within $[ \frac{\pi }{2},\frac{3\pi }{2}]$ are isomorphic. In particular, the narrow beam in direction $\varphi$ is equivalent to that in direction $\pi-\varphi$, i.e.,
\begin{equation}
{{\bf{a}}_{{N_a}}}(\varphi)= {{\bf{a}}_{{N_a}}}(\pi-\varphi ).\label{lem1}
\end{equation}
\end{lemma}
\begin{IEEEproof}
Following the convention, we define the angle perpendicular to the array plane as 0 and accordingly the front range (FR) is $[ - \frac{\pi }{2},\frac{\pi }{2}]$.  It is easy to verify (\ref{lem1}) due to the fact $\sin (\varphi ) = \sin (\pi  - \varphi )$.
\end{IEEEproof}
\begin{figure}[t]
\center
\includegraphics[width=2.8in]{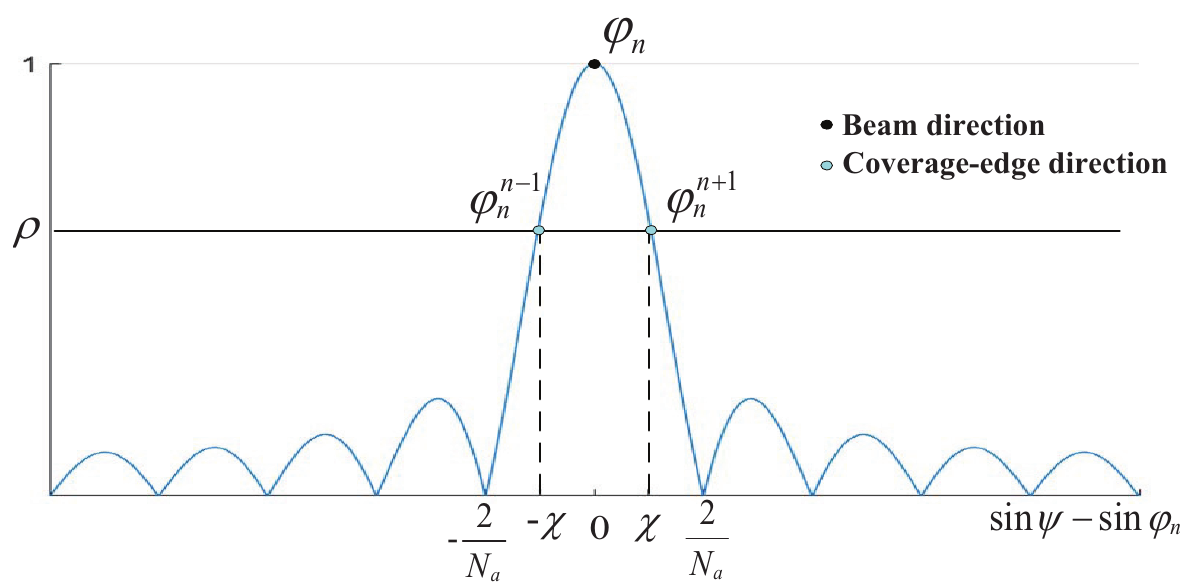}
\caption{Normalized $i$th beam power distribution in different directions.}\label{figsin}
\vspace{-12pt}
\end{figure}

For simplicity, we use the notation $\varphi _n$ for a pair of front-back symmetric beams. Next, we design the distribution of the $N$ narrow beams, i.e, design the directions of $\{\varphi_n\}_{n=1}^N$, which are different from the uniform distribution as proposed in \cite{track2}. To this end, we first define the beam coverage of a narrow beam ${{\bf{a}}_{{N_a}}}(\varphi )$ as \cite{track3}
\begin{equation}\label{cv}
{\mathcal{CV}}\left( {{{\bf{a}}_{{N_a}}}(\varphi )} \right) = \left\{ {\psi \left| {\; {A\left( {{{\bf{a}}_{{N_a}}}(\varphi ),\psi } \right)}  \ge \rho } \right.} \right\}
\end{equation}
where $A\left( {{{\bf{a}}_{{N_a}}}(\varphi ),\psi } \right)$ is the normalized detected beam gain of ${{\bf{a}}_{{N_a}}}(\varphi )$ in the direction of $\psi$, i.e.,
\begin{align}
A\left( {{{\bf{a}}_{{N_a}}}(\varphi ),\psi } \right) &= \left| {{{\bf{a}}_{{N_a}}}{{(\varphi )}^H}{{\bf{a}}_{{N_a}}}(\psi )} \right|\label{bpwer}\\
 &= \left| {\frac{1}{{{N_a}}}\sum\limits_{n = 1}^{{N_a}} {{e^{jk{d_a}(n - 1)[\sin \psi  - \sin \varphi ]}}} } \right| \le 1.\notag
\end{align}
As shown in Fig. \ref{figsin}, we plot the distribution of $A\left( {{{\bf{a}}_{{N_a}}}(\varphi_n ),\psi } \right)$ versus $\sin \psi  - \sin \varphi_n$. It is observed that the gain can be maximally achieved in the direction of $\varphi_n$ and decreases in other directions. In this regard, $\rho$ in (\ref{cv}) denotes a gain threshold, which determines the beam coverage, i.e., the direction range within which the normalized beam gain is no smaller than $\rho$. Therefore, we refer to $\rho$ as coverage-edge gain.
To design the distribution of the $N$ beams $\{{{\bf{a}}_{N_t}}(\varphi_n)\}_{n=1}^N$, we use $\varphi _n$ and $\varphi _n^{n\pm 1}$ to represent the central direction and the coverage-edge directions of the $n$th beam, respectively. The length of coverage of the $n$th beam is defined as the beamwidth $\Delta \varphi_n$ with $\Delta \varphi_n=\varphi _n^{n+1}-\varphi _n^{n-1}$. For half-wavelength antenna spacing, we have the following Proposition and Corollary concerning the distribution of the beams with the same $\rho$ and with the same beamwidth (i.e., $\Delta \varphi_n$ is the same for all $n$), respectively.
\begin{figure}[t]
\center
\includegraphics[width=2.8in]{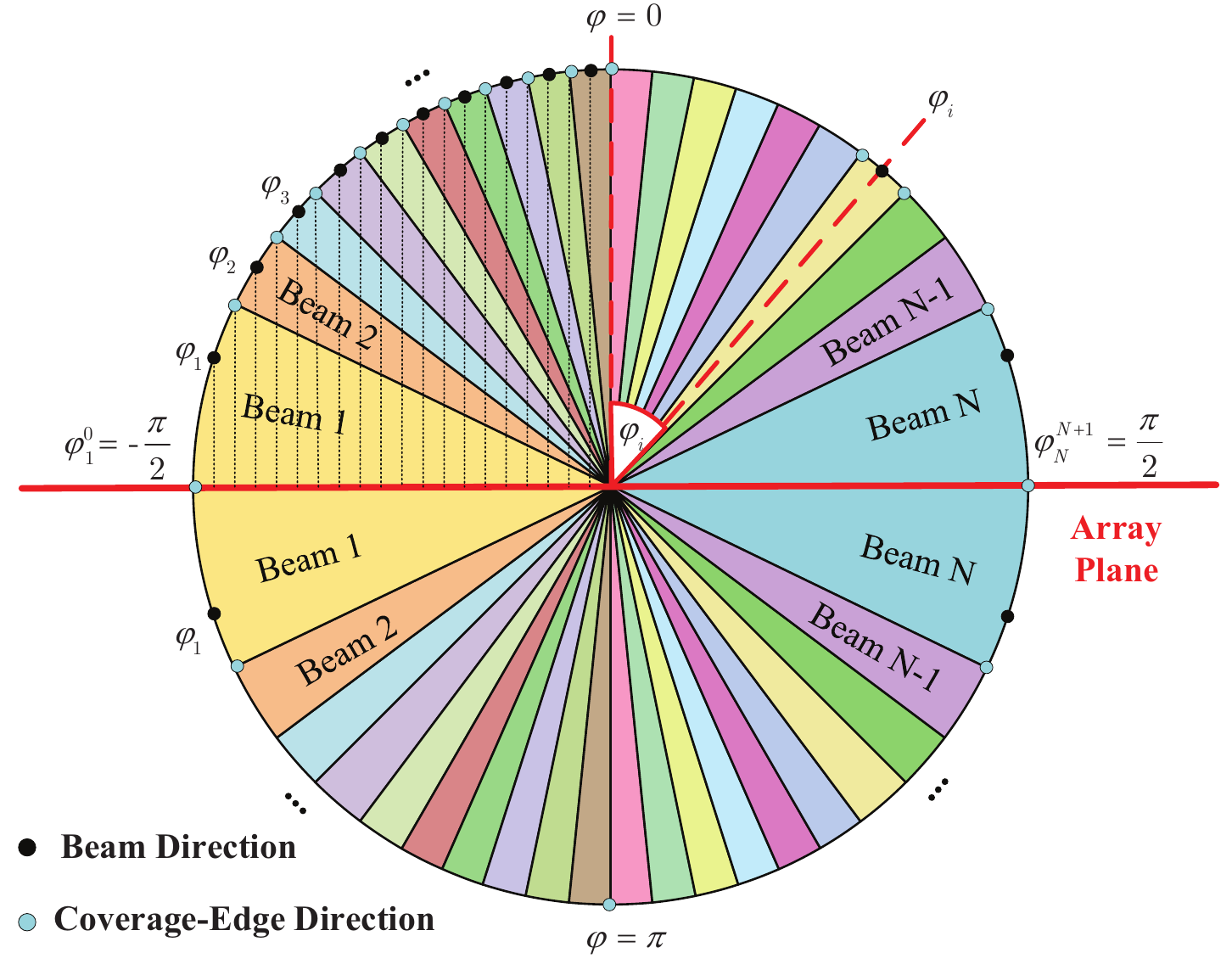}
\center
\caption{Narrow-beam patterns in different directions with same $\rho$.}\label{figpat}
\vspace{-8pt}
\end{figure}
\begin{proposition}
If $N$ narrow beams $\{{\bf{a}}_{N_a}(\varphi_n)\}_{n=1}^N$ cover the whole space with the same coverage-edge gain, we have
\begin{equation}
\sin \varphi _n^{n + 1} - \sin {\varphi _n} = \sin {\varphi _n} - \sin \varphi _n^{n - 1}= \frac{1}{N},
\end{equation}
and the coverage-edge gain $\rho$ is
\begin{equation}
\rho  = \frac{{\sin [({N_a}\pi)/2N]}}{{{N_a}\sin [\pi/2N]}},
\end{equation}
which is monotonically increasing with $N$. As such, the directions of $\{ {\varphi _n}\} _{n = 1}^{N}$ are given by
\begin{equation}\label{direc}
{\varphi _n}\! =\! \left\{ {\begin{split}
{\arcsin \left[ {\frac{{2n \!-\! 1}}{N} \!-\! 1} \right],\;{\varphi_n \in \rm{FR}}\;}\\
{\pi  \!-\! \arcsin \left[ {\frac{{2n \!-\! 1}}{N} \!-\! 1} \right],\;{\rm{otherwise}}}
\end{split}} \right., n\!=\!1,2,...,N.
\end{equation}
\end{proposition}

To better illustrate (\ref{direc}), Fig. \ref{figpat} shows the narrow beams covering the whole space with the same $\rho$. It is observed that the beams have different beamwidth $\Delta \varphi$. However, Proposition 1 implies that these beams have the same spacing in the sine space\footnote{We use the dotted lines in the upper left corner of Fig. 5 to discern the sine spacing, and it is clear that they have the same spacing.}, i.e., $\sin(\varphi _i^{i+1})-\sin(\varphi _i^{i-1})=2/N$. 
\begin{corollary}
If the $N$ narrow beams $\{{\bf{a}}_{N_a}(\varphi_n)\}_{n=1}^N$ cover the whole space with the same beamwidth, i.e.,  $\Delta \varphi  = \pi/N$, the beam directions of $\{ {\varphi _n}\} _{n = 1}^{N}$ are given by
\begin{equation}\label{nonui}
 {{\varphi _n} = \left\{ {\begin{split}
{ - \frac{\pi }{2} + \frac{{(2n - 1)\pi }}{{2N}},\;\;{\varphi_n \in \rm{FR}}\;\;}\\
{\;\;\frac{{3\pi }}{2} - \frac{{(2n - 1)\pi }}{{2N}},\;\;{\rm{otherwise}}}
\end{split}} \right.}
\end{equation}
for $i=1,2,...,N$ with 
\begin{equation}
{\cal C}{\cal V}({{\bf{a}}_{{N_a}}}({\varphi _n})) \!=\! \left\{ {\begin{split}
{\left[ { - \frac{\pi }{2} \!+\! \frac{{(n - 1)\pi }}{N}, \!-\! \frac{\pi }{2} \!+\! \frac{{n\pi }}{N}} \right],\;{\varphi_n \in \rm{FR}}\;}\\
{\left[ {\frac{{3\pi }}{2} \!-\! \frac{{n\pi }}{N},\;\frac{{3\pi }}{2} \!-\! \frac{{(n - 1)\pi }}{N}} \right],\;{\rm{otherwise}}}
\end{split}} \right..\;
\end{equation}
In this case, the coverage-edge gain $\rho$ is variable and lower bounded by the following function of $\Delta \varphi$ for all beams, i.e.,
\begin{equation}
\rho  > \frac{{\sin [(\Delta \varphi {N_a}\pi)/4]}}{{{N_a}\sin [(\Delta \varphi \pi)/4]}}.
\end{equation}
\end{corollary}
\begin{IEEEproof}
Proofs of Proposition 1 and Corollary 1 are relegated to Appendix A.
\end{IEEEproof}
\begin{figure}[t]
\center
\includegraphics[width=2.6in]{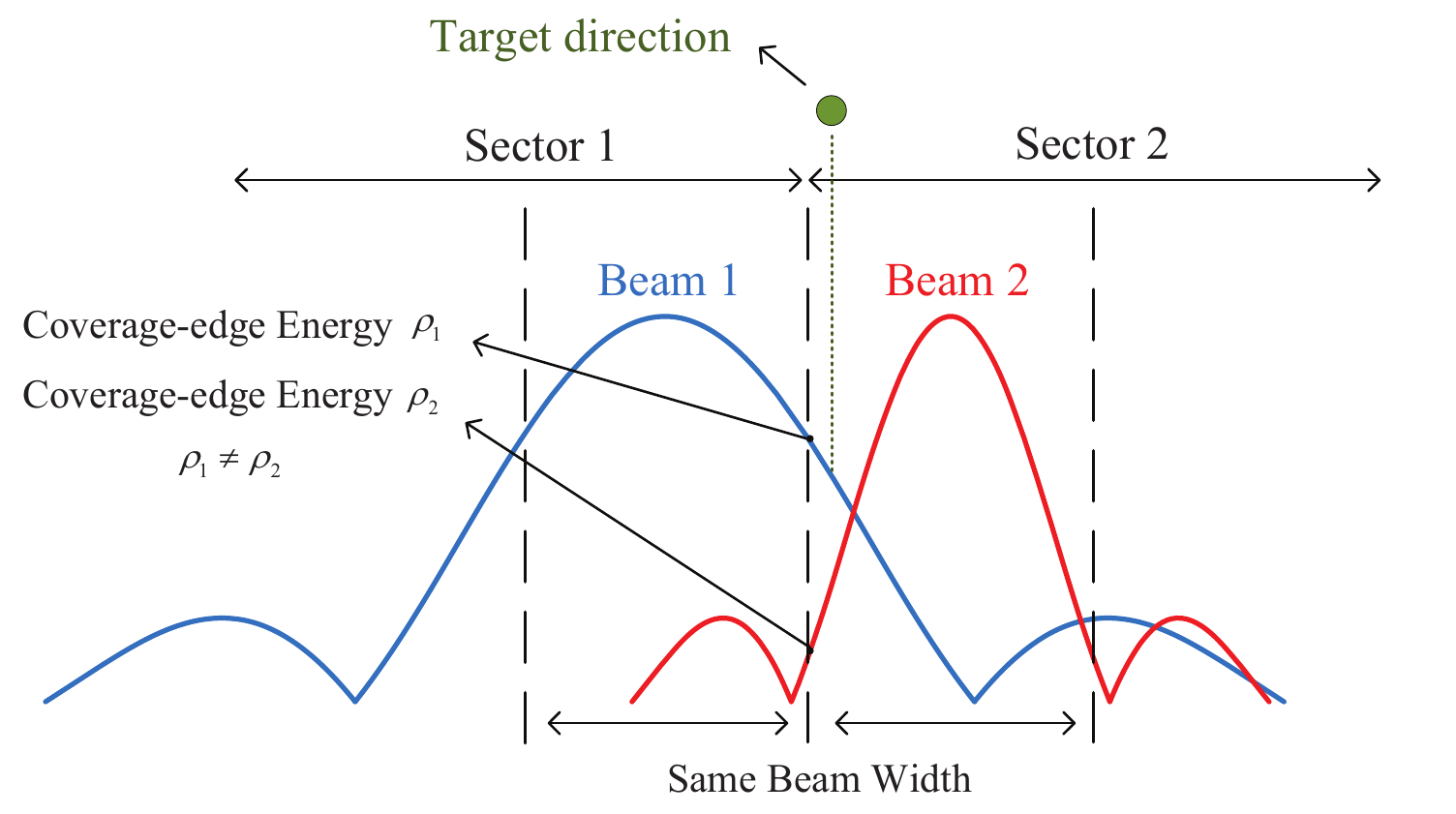}
\caption{Two beams with the same beamwidth but with different $\rho$.}\label{mis}
\vspace{-12pt}
\end{figure}

Note that\cite{track2} considered the uniformly distributed beams (i.e., beams with the same beamwidth) as in Corollary 1, we propose to apply the beams with the same $\rho$ as in Proposition 1 due to the following reasons:
\begin{itemize}
\item A misalignment issue may happen if the uniformly distributed beams are adopted in the hierarchical search (which will be specified in Section IV-B). For example, as seen from Fig. \ref{mis}, in the hierarchical search, the target direction will be discerned as in sector 2 by using wide beams. Then, only narrow beams in sector 2 are needed for fine search. As such, beam 2 will be selected as a solution while beam 1 is the best one. 
\item Since the beams are narrower around direction $0$ (perpendicular to the array plane) and wider around direction $\pm \frac{\pi }{2}$, the uniform beams are not close enough around direction $0$ and there is no worst-case performance guarantee. In contrast, the proposed beams are distributed tightly around direction $0$ but loosely around direction $\pm \frac{\pi }{2}$, which ensures that the normalized detection power is larger than $\rho$. 
\end{itemize}
The advantages of the proposed beams over the uniformly distributed beams in \cite{track2} will also be more clearly shown in Fig. \ref{dtrate} and \ref{n67} in Section VI.


\subsection{Beam Training for Reflecting Links}\label{IRSt}
Despite the angles for BS-user$k$ path can be estimated by steering beams as previously mentioned, this method cannot be directly applied to the BS-IRS$l_k^m$ path and IRS$l_k^m$-user$k$ path since the IRS cannot transmit/receive beams by itself.

As such, we propose an exhaustive beam training method for the BS-IRS$l_k^m$-user$k$ cascaded path.  Assume that all AoDs/AoAs are taken from the $N$ points given in (\ref{direc}), the cascaded path must be one of  the $N\times N^2 \times N=N^4$ paths. In our prior work [1], we have shown how to determine the optimal combination manner for any single path. Therefore, we could enumerate and compare the communication performance by all the $N^4$ combinations and find the optimal one, then obtain the corresponding angles $\varphi _{B,M}^{k,m},\varphi _{R,M}^{k,m}, \varphi _{R,N}^{k,m},$ and $\varphi _{U,N}^{k,m}$ for the BS-IRS$l_k^m$-user$k$ cascaded path.

\section{Low-Complexity Beam Training Method}\label{channeles}
Though the path angles can be optimally found via the exhaustive beam training approach, it is impractical to implement it in massive MIMO systems due to the unaffordable training overhead, especially for large $N$. Based on the narrow beams given in Proposition 1, in this section, we propose a low-complexity beam training scheme that combines partial search at IRS and a ternary-tree hierarchical search at BS and users. 
\subsection{Partial Search for IRS-PSs}\label{PSO}
Based on ({\ref{sepchan}), the BS-IRS$l_k^m$-user$k$ cascaded path (normalized by the antenna gains) can be expressed as 
\begin{small}
\begin{align}\label{22}
&\frac{{{\bf{N}}_k^m{{\bf{\Theta }}_k^m}{{\bf{M}}_k^m}}}{{a(f,d_{k,m}^N)a(f,d_{k,m}^M)}}\\
&= {\bf{a}}_{{N_u}}^U\left( {\varphi _{U,N}^{k,m}} \right)\underbrace {{\bf{a}}_{{N_r}}^B{{\left( {\varphi _{R,N}^{k,m}} \right)}^H}{{\bf{\Theta }}_k^m}{\bf{a}}_{{N_r}}^R\left( {\varphi _{R,M}^{k,m}} \right)}_{{\text{effective\;gain }}}{\bf{a}}_{{N_t}}^B{\left( {\varphi _{B,M}^{k,m}} \right)^H},\notag
\end{align}
\end{small}and the optimal PSs of IRS$l_k^m$ should maximize the effective gain in (\ref{22}), which can be rewritten as
\begin{equation}\label{apaa}
\begin{split}
&\left| {{\bf{a}}_{{N_r}}^B{{\left( {\varphi _{R,N}^{k,m}} \right)}^H}{{\bf{\Theta }}_k^m}{\bf{a}}_{{N_r}}^R\left( {\varphi _{R,M}^{k,m}} \right)} \right|\\
&=\frac{\beta }{{{N_r}}}\left| {\sum\limits_{n = 1}^{{N_r}} {{e^{j[k{d_a}(n - 1)(\sin \varphi _{R,M}^{k,m} - \sin \varphi _{R,N}^{k,m}) + {\theta _n}]}}} } \right| \le \beta.
\end{split}
\end{equation}
Thus, an optimal IRS-PS solution can be expressed as 
\begin{align}\label{apa}
&{\bf{\Theta }}_k^{m,\rm{opt}} = {\rm{diag}}(\beta {e^{j\theta _1^{}}},\beta {e^{j\theta _2^{}}}, \cdots ,\beta {e^{j\theta _{{N_r}}^{}}}),\\
&{\theta _n} = k{d_a}(n - 1)(\underbrace {\sin \varphi _{R,N}^{k,m} - \sin \varphi _{R,M}^{k,m}}_{\triangleq \;\Delta _{k,m} }),\;\;n = 1,2,...,{N_r},\notag
\end{align}
where $\Delta _{k,m}$ is the angle difference in the sine space.
\begin{proposition}
The set of optimal IRS-PSs solutions in (\ref{apa}) is smaller than that of the considered paths by exhaustive search. More specifically, there are only $2N-1$ candidate IRS-PS solutions versus $N^2$ possible paths with different AoD/AoA combinations. 
\end{proposition}
\begin{IEEEproof}
Given the AoDs/AoAs in  (\ref{direc}), the path angles are taken from a uniform grid of $N$ points in the sine space with $\sin \varphi  \in \{  - 1 + \frac{1}{N}, - 1 + \frac{3}{N},...,\frac{1}{N} - 1\} $. As $|\sin {\varphi _{i+1}} - \sin \varphi _i| = \frac{2}{N}$, we have ${\Delta _{k,m}} \in \{  - 2 + \frac{2}{N}, - 2 + \frac{4}{N},...,2 - \frac{2}{N}\}$ with $2N-1$ candidate solutions.
\end{IEEEproof}

It follows from Proposition 2 that there is no need to search the exact path angles at IRS itself (i.e., $\varphi _{R,M}^{k,m}, \varphi _{R,N}^{k,m}$ in Fig. \ref{fig3n}) but only need to search $\Delta _{k,m}$ in (\ref{apa}), which helps to reduce the complexity.
\subsection{Ternary-Tree Hierarchical Search}\label{TTH}
\begin{figure}[t]
\center
\includegraphics[width=3.0in]{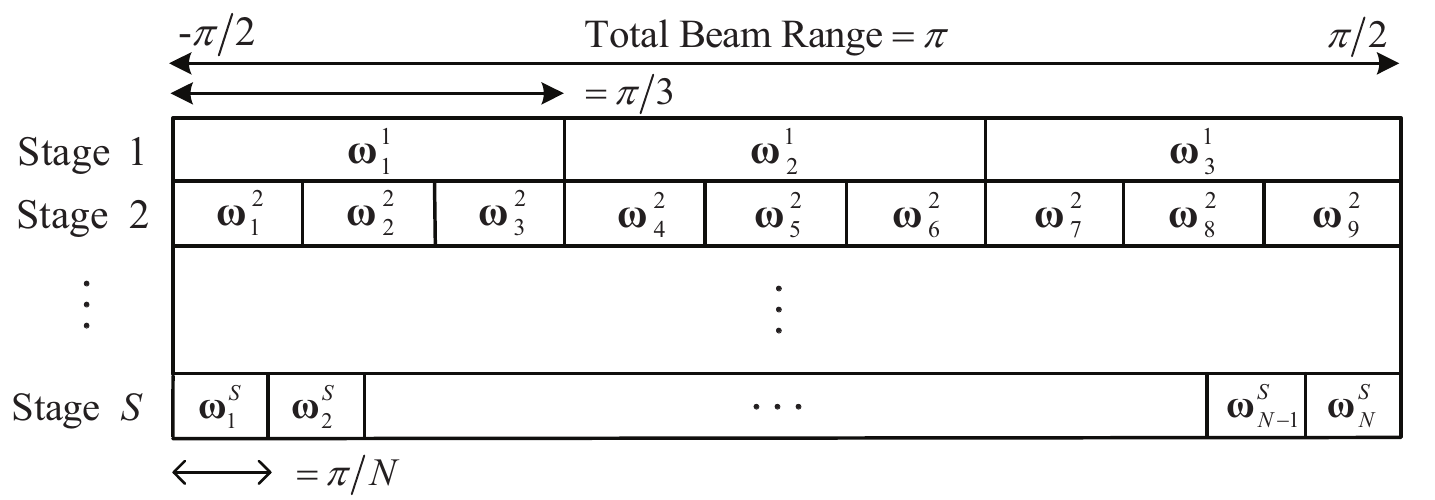}
\caption{Beam coverage structure of the ternary-tree hierarchical codebook.}\label{fig3t}
\vspace{-12pt}
\end{figure}
 To find the path angles at BS and the users, we propose a ternary-tree hierarchical search that realizes the beam search via multiple stages with wide beams and narrow beams, rather than exhaustively searching over all narrow-beam pairs.  Let ${\bm{\omega }}_n^s$ denote the $n$th beam vector in the $s$th stage, Fig. \ref{fig3t}  shows the beam coverage structure of a ternary-tree codebook, where $S=\log _3N$ is the number of the bottom stage. In particular, the codewords $\{{\bm{\omega }}_n^{{S}}\}_{n=1}^N$ in stage $S$ is exactly the narrow-beam codewords proposed for exhaustive search, i.e,
\begin{equation}\label{phi}
{\bm{\omega }}_n^{{S}} = {{\bf{a}}_{{N_a}}}({\varphi _n}),\;
{\varphi _n} \;{\rm{is\; given\; by \;(\ref{direc})}}.
\end{equation}

While hierarchical searching schemes are commonly adopted as the binary-tree search \cite{track1,track2,track3,track4}, we first establish that the ternary-tree search is actually more efficient than the binary-tree search, as shown in the following Proposition.
  \begin{proposition}
Ternary-tree search has the lowest searching complexity among all tree search schemes.
 \end{proposition}
 \begin{IEEEproof}
Assume that $N$ narrow beams are implemented in the bottom stage. In $M$-ary tree search scheme, the range covered by a beam in the current stage is $M$ times as much as that in the next stage. As such, the total number of stages is $\log _M{N}$ and the search time (number of tests) taken to find the best narrow beam is given by
 	\begin{equation}
	T(M,N)=M\log_{M}N.
\end{equation}
 Assuming $M$ is continuous, we can find $M$ that minimizes $T(M,N)$ by taking the derivative with respect to $M$,
\begin{equation}
\frac{{\partial T(M,N)}}{{\partial M}}=\frac{{(\ln M - 1)\ln N}}{{{{\left( {\ln M} \right)}^{\rm{2}}}}} = 0\quad \Rightarrow M = e.
\end{equation}
For discrete $M$, we compare $T(M,N)$ on two adjacent integers around $e$. Define ${q}(N) = T(2,N) - T(3,N) = 2\log _{2}N - 3\log _{3}N$. It follows that ${q}(N)$ increases monotonically with $N$ since ${\rm{d}}{q}(N){\rm{/d}}N = (2/\ln 2 - 3/\ln 3)/N > 0$. Due to ${q}(1)>0$, we have $q(N)>0$ holds true for all discrete $N\ge 1$. Thus, ternary-tree search leads to the minimum search time among $M$-ary tree search schemes. To verify this, we plot the search time of different $M$-ary tree schemes in Fig. \ref{mt}.
 \end{IEEEproof}
 \begin{figure}[t]
\center
\includegraphics[width=3.2in]{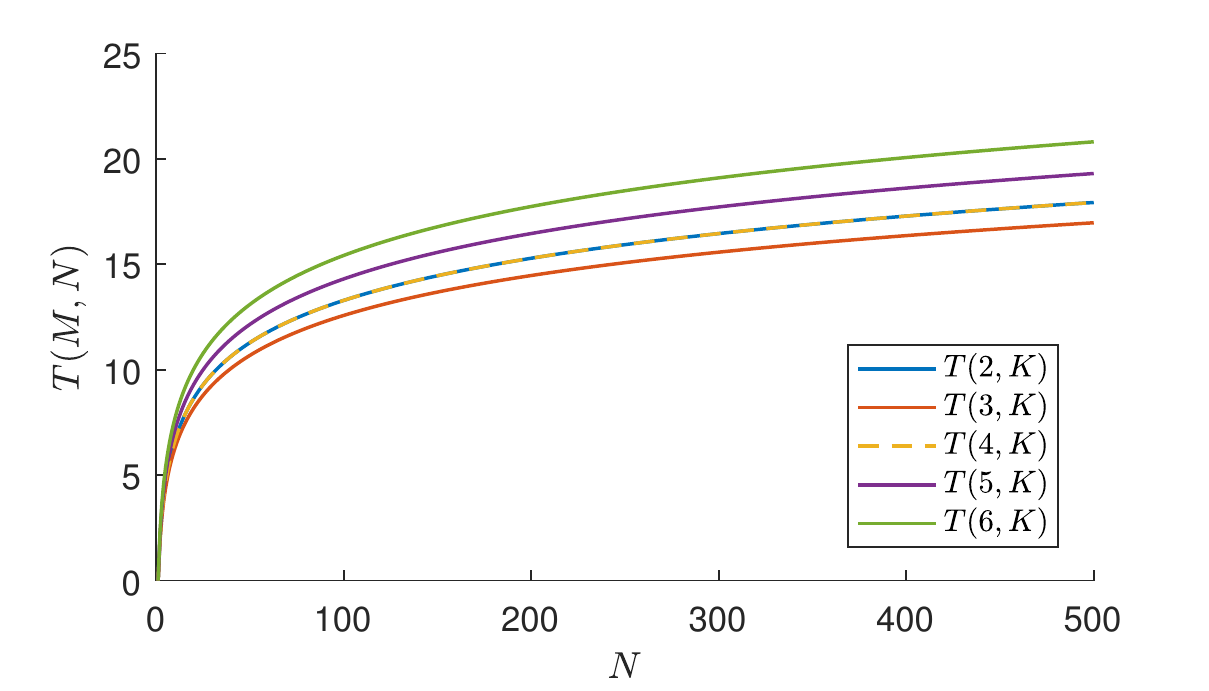}
\caption{The search time of $M$-ary tree scheme with $M=2,3,4,5,6$.}\label{mt}
\vspace{-12pt}
\end{figure}
\begin{figure}[t]
\center
\includegraphics[width=3.2in]{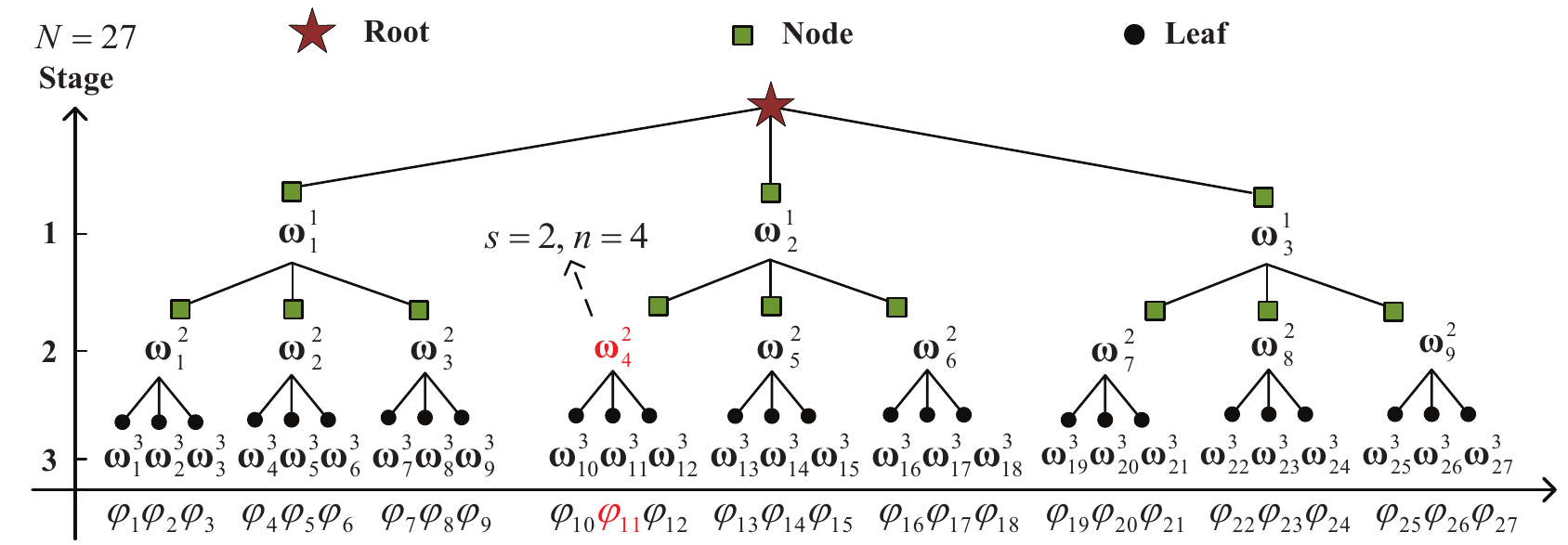}
\caption{Beam candidates of ternary-tree hierarchical scheme when $N=27$.}\label{27}
\vspace{-12pt}
\end{figure}
Next, we design two novel training codebooks, namely, TD codebook and PSD codebook, for realizing all beams used in the ternary-tree search.  As shown in Fig. \ref{27},  each wide beam covers three descendant beams in the next stage. 
 \subsubsection{TD Codebook}
The narrow-beam codewords $\{{\bm{\omega }}_n^{{S}}\}_{n=1}^N$ in the bottom stage are given in (\ref{phi}). To design the wide-beam codewords in the upper stages of the ternary-tree structure, i.e., $\{{\bm{\omega }}_n^s\}_{s=1}^{S-1}\in {\mathcal{C}^{{N_a} \times 1}}$,  we expect that 
\begin{equation}\label{cb}
{({\bm{\omega }}_i^{S})^H}{\bm{\omega }}_n^s = \left\{ {\begin{split}
&{1,\;\;\;{\bm{\omega }}_i^{S}{\rm{\;is\;a\;descendant}}\;{\rm{of}}\;{\bm{\omega }}_n^s}\\
&{0,\;\;\;\;\;\;\;\;\;\quad{\rm{otherwise}}}
\end{split}} \right.
\end{equation}
holds true for all $i=1,2,...,N$.  Let ${{\bm{\Xi }}_s} = [{\bm{\omega }}_1^s\;{\bm{\omega }}_2^s\;...\;{\bm{\omega }}_{{3^s}}^s]$ represent the wide beams in stage $s$. Then, we can rewrite (\ref{cb}) in a more compact form as
\begin{equation}\label{com}
{[{\bm{\omega }}_1^{{S}}\;{\bm{\omega }}_2^{{S}}\;...\;{\bm{\omega }}_{N}^{{S}}]^H}{\bm{\Xi }}_s = {{\bf{L}}^H}{\bf{\Xi }_s} = {{\bf{D}}_s},
\end{equation}
where ${{\bf{D}}_s}$ is an $N \times 3^s$ matrix. The $i$th column of ${{\bf{D}}_s}$ has an element of 1 in the rows
$
\{(i - 1){3^{{S} - s}} + n, |\;n = 1,2,...,{3^{{S} - s}}\}
$
and an element of 0 in other rows. For ease of understanding, as shown in Fig. \ref{d12}, we provide an example of the explicit ${{\bf{D}}_s}$ in (\ref{com}) when $N=27$. As we can seen, each column of ${\bf{D}}_s$ can reflect the beam coverage of a codeword. As a result, the $n$th codeword in stage $s$ can be obtained as 
\begin{figure}[t]
\center
\includegraphics[width=2.6in]{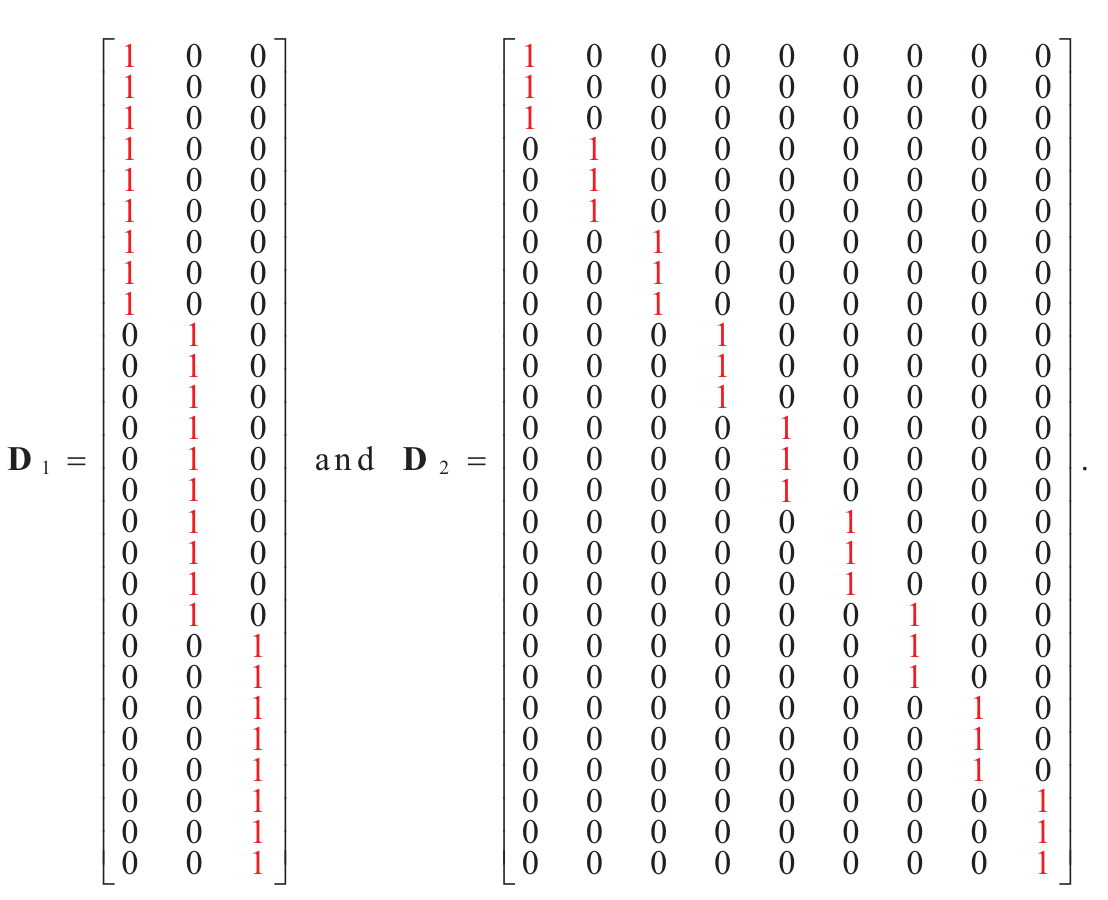}
\caption{${\bf{D}}_1$ and ${\bf{D}}_2$ in (\ref{com}) when $N=27$.}\label{d12}
\vspace{-12pt}
\end{figure}

\begin{equation}\label{bems}
{\bm{\omega }}_n^s = \;{({\bf{L}}{{\bf{L}}^H})^{ - 1}}{\bf{L}}{{\bf{D}}_S}(:,n).
\end{equation}
With an HB architecture, this codeword is realized by ${\bm{\omega }}_n^s = {\bf{F}}_{RF}{\bf{f}}_B^{s,n}s$, for which ${\bf{F}}_{RF}$ and ${\bf{f}}_B^{s,n}s$ can be determined by solving a classical minimum Euclidean distance problem:
\begin{subequations}\label{omp} 
\begin{align}
\{ {{\bf{F}}_{RF}},{\bf{f}}_B^{s,n}\}  =& \arg \min {\left\| {{\bm{\omega }}_n^s - {{\bf{F}}_{RF}}{\bf{f}}_B^{s,n}} \right\|_F}\\
&\;\;{\rm{s}}{\rm{.t}}{\rm{.}}\;\;\;\;\;\;\left\| {{{\bf{F}}_{RF}}{\bf{f}}_B^{s,n}} \right\|{\rm{ = 1}},\;\\
&\;\;\;\;\;\;\;\qquad{{\bf{F}}_{RF}}(:,i)  \in \mathcal{F}_{\rm{RF}},\;\;\forall i.\label{phas}
\end{align}
\end{subequations}
Problem (\ref{omp}) has been well studied and the orthogonal matching pursuit (OMP) algorithm can be applied to resolve it\cite{eud4}.

It is worth noting that by using the TD codebook, each wide beam realization requires all RF chains. In the following, we further propose a PSD codebook, whereby only one RF chain is needed for each wide beam realization.  Thus, the overall searching time can be reduced by simultaneously testing multiple beams. Despite that the PSD codebook yields lower search complexity, we should mention that the TD codebook is still appealing due to its stronger anti-noise capability, as will be shown in Section \ref{CDB}.

\subsubsection{PSD Codebook}
Still, the narrow-beam codewords $\{{\bm{\omega }}_n^{{S}}\}_{n=1}^N$ in the bottom stage are given in (\ref{phi}). 
The key of the PSD codebook is to let each RF chain control a subset of antenna elements by deactivating some PSs, so as to realize wide beam via less active antenna elemtents. Specifically, each RF chain occupies ${3^s}$ adjacent antenna elements to transmit/receive beam in stage $s=1,2,...,s_{\rm{max}}$, with $s_{\rm{max}}=\left\lfloor {\log _3{{N_a}}} \right\rfloor$. Whereas in stage $s=s_{\rm{max}}+1,...,S$, all $N_a$ antennas are used to transmit/receive beam. Thus, the $n$th beam codeword in stage $s$ can be expressed as 
\begin{equation}\label{c3}
{\bm{\omega }}_n^s = \left\{ {\begin{split}
&{{\bf{a}}_{{3^s}}({\varphi _{i(N,n,s)}}),\;\;\;\;\;\;s = 1,2,...,{s_{\max }}\;\;\;}\\
&{{\bf{a}}_{N_a}({\varphi _{i(N,n,s)}}),\;\;s = {s_{\max }} + 1,...,S}
\end{split}} \right.,
\end{equation}
and the angle index is given by
\begin{equation}\label{c3i}
i(N,n,s) = {\frac{{N \cdot {3^{{\rm{ - }}s}}(2n - 1) + 1}}{2}},
\end{equation}
where the definition of $\varphi _i$ is the same as that in (\ref{phi}). To illustrate the angle index in  (\ref{c3i}), we present an angle map in Fig. \ref{27} with setting $N=27$, where the angle of each beam is depicted vertically below it. For example, the angle index below the beam $\bm{\omega }_4^2$ in Fig. \ref{27} is $11$ as the red one. It should also follows from (\ref{c3i}) that $i(27,4,2)=11$.

With the PSD codebook, the coverage of each wide beam ${\bm{\omega }}_n^s$ ($s\neq S$) can be similarly defined as in (\ref{cv}) and we have the following proposition.
\begin{proposition}
Consider the PSD codebook. By setting the coverage-edge gain $\rho$ of each beam in stage $s$ as 
\begin{equation}\label{p4}
\rho (s) = \left\{ {\begin{split}
&\frac{1}{{{3^s}\sin \left( {\frac{\pi }{{2 \cdot {3^s}}}} \right)}},\;\;\;\;\;s = 1,...,{s_{\max }}\\
&\frac{{\sin \left( {\frac{{{N_a}\pi }}{{{3^s}}}} \right)}}{{{N_a}\sin \left( {\frac{\pi }{{2 \cdot {3^s}}}} \right)}},\;\;s = {s_{\max }} + 1,...,S
\end{split}} \right., 
\end{equation}
it follows that each beam in stage $s$ ($s\neq S$) can cover three beams in stage $s+1$, i.e., 
\begin{equation}
{\cal C}{\cal V}({\bm{\omega }}_n^s) = {\cal C}{\cal V}({\bm{\omega }}_{3n - 2}^{s + 1}) \cup {\cal C}{\cal V}({\bm{\omega }}_{3n - 1}^{s + 1}) \cup {\cal C}{\cal V}({\bm{\omega }}_{3n}^{s + 1}).
\end{equation}.
\end{proposition}
\begin{IEEEproof}
The proof is relegated to Appendix B.
\end{IEEEproof}

While the bottom-stage narrow beams are shown to cover the whole space by Proposition 1, Proposition 4 implies that by setting a proper $\rho$ for each upper stage,  the wide beams in the same stage can still cover the whole space since each beam exactly covers three beams in the next stage. 
\subsection{Cooperative Beam Training Procedure}\label{ccep}
\begin{figure}[t]
\center
\includegraphics[width=3.0in]{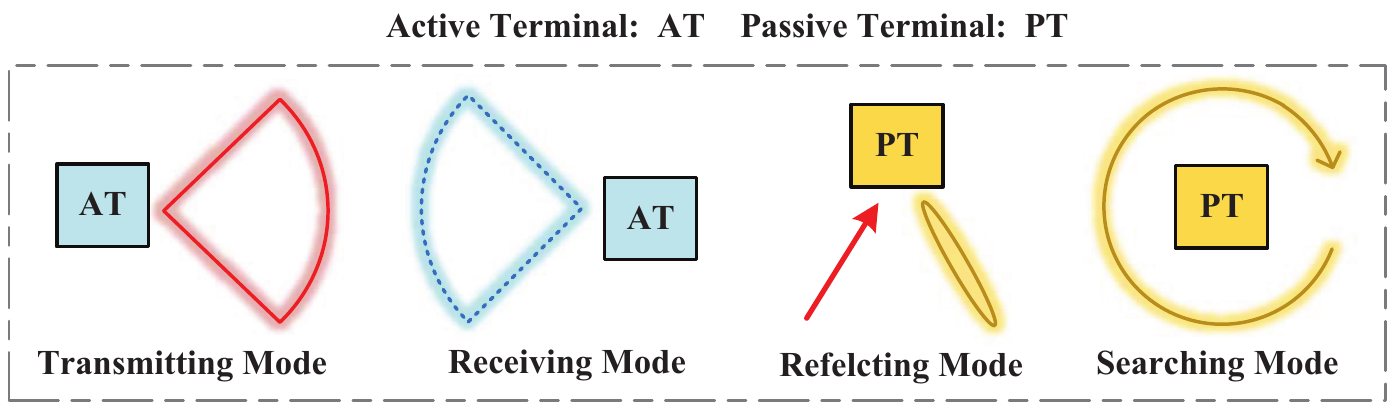}
\caption{Beam modes for the active terminal (BS or user) and that for the passive terminal (IRS)  in the proposed beam training procedure, respectively.}\label{figmode}
\vspace{-12pt}
\end{figure}
Based on the partial search approach and the ternary-tree codebooks, we develop a cooperative training procedure in this subsection.  For ease of exposition, we first define four modes as below and also illustrated in Fig. \ref{figmode}.
\begin{itemize}
\item \emph{Transmitting Mode:} \\
We use solid red line to represent the transmit beams. In this mode, the transmitter uses ${\bm{\omega }}$ as the precoder to form beam in a certain coverage range. If the AoD of a propagation path ${\bf{H}}$ is within this coverage range, we have $||{\bf{H}}{\bm{\omega }}||_2\gg 0$; otherwise, we have $||{\bf{H}}{\bm{\omega }}||_2 \to 0$.

\item \emph{Receiving Mode:} \\
The solid-broken blue line denotes the receive beams. In this mode, the receiver uses ${\bm{\omega }}$ as the combiner to detect the signal in a certain range, so as to determine the existence of path. If the AoA of a propagation path ${\bf{H}}$ is within this beam range, we have $||{\bm{\omega}}^H{\bf{H}}||_2\gg 0$; otherwise,  we have $||{\bm{\omega }}^H{\bf{H}}||_2\to 0$.
\item \emph{Reflecting Mode:} \\
We use solid red arrow and solid yellow line to represent the incoming signal and the reflected signal, respectively.
In this mode, the IRS-PSs are set in the following form:
\begin{equation}
\begin{split}
&{\bf{\Theta }}_{\rm{ref}}(x) = {\rm{diag}}(\beta {e^{j\theta _1^{}}},\beta {e^{j\theta _2^{}}}, \cdots ,\beta {e^{j\theta _{{N_r}}^{}}}),\\
&{\theta _n} = k{d_a}(n - 1)x,\;\;\;n = 1,...,{N_r}.
\end{split}
\end{equation}
It is easy to verify that for any incoming narrow-beam vector ${{\bf{a}}_{{N_r}}}$, the reflected signal ${\bf{\Theta }}_{\rm{ref}}(x){{\bf{a}}_{{N_r}}}$ still follows the narrow-beam structure as in (\ref{ula}).

\item \emph{Searching Mode:} \\
The yellow round arrow represents that the IRS is applying different predefined codewords successively the in reflecting mode.
In this mode, we need to search $\Delta _{k,m}$ in (\ref{apa}).  Specifically, the IRS applies ${\bf{\Theta }}_{\rm{ref}}(- 2 + \frac{{2i}}{N})$, $i=1,2,...,2N-1$, one by one in $2N-1$ time slots.
\end{itemize}

To obtain the required channel information, i.e., $\Delta _{k,m}$ and path angles at the BS and user$k$ (see Fig. \ref{fig3n}), three phases are developed to achieve different groups of measurements in a cooperative manner as stated below. 
\begin{figure}[t]
\center
\includegraphics[width=3.4in]{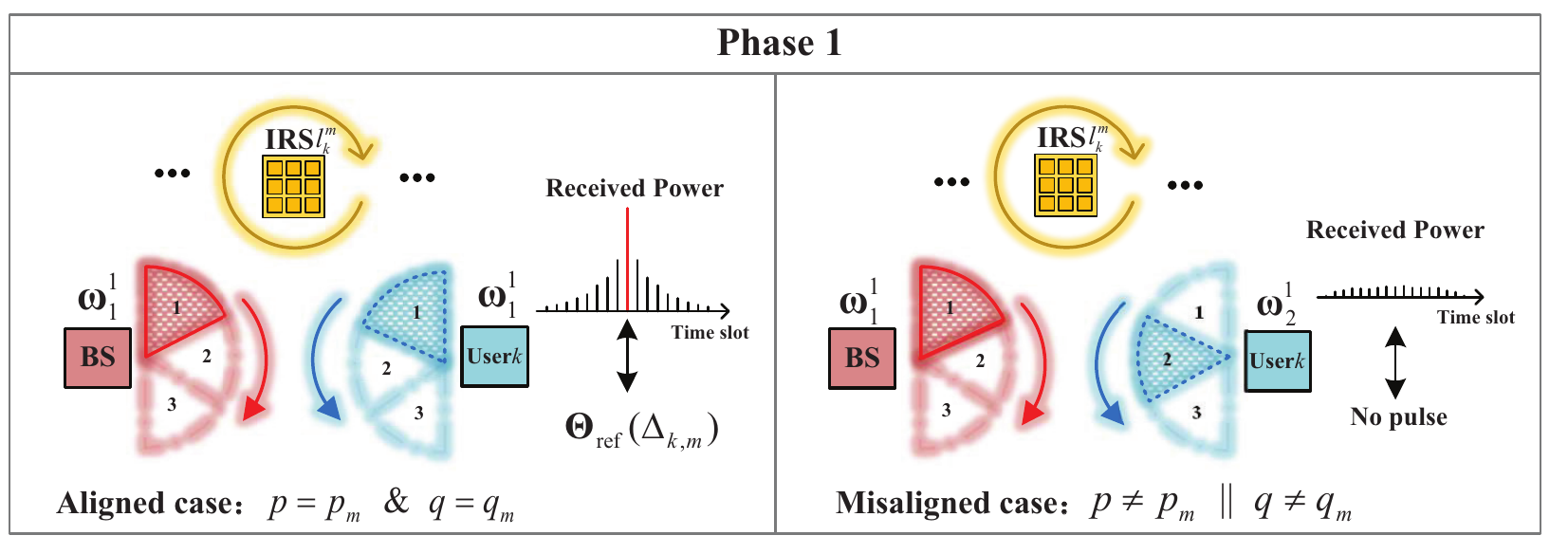}
\caption{Phase 1 of the cooperative beam training procedure.}\label{figp1}
\vspace{-12pt}
\end{figure}
\subsubsection{Phase 1 (see Fig. \ref{figp1})}
In Phase 1, we aim to obtain $\Delta _{k,m}$ in (\ref{apa}).  To this end,  we test $3\times3$ wide beams in stage 1 in $9$ successive intervals with BS using $\{{\bm{\omega }}_p^1\}_{p=1,2,3}$ in the transmitting mode and user$k$ using $\{{\bm{\omega }}_q^1\}_{q=1,2,3}$ in the receiving mode. The resulting signal can be expressed as
\begin{equation}\label{9p}
{y_{q,p}} = \sqrt P {({\bm{\omega }}_q^1)^H}{{\bf{H}}_k}{\bm{\omega }}_p^1s + n,
\end{equation}
where $s=1$ is the normalized pilot signal. In each interval, all IRSs apply the searching mode simultaneously in $2N-1$ time slots. For IRS$l_k^m$, there is only one beam pair ($p_m \times q_m$) that covers both the BS-IRS$l_k^m$ link and the IRS$l_k^m$-user$k$ link (see aligned case in Fig. \ref{figp1}), i.e.,
\begin{equation}
{({\bm{\omega }}_{{q_m}}^1)^H}{\bf{H}}_{k,l}^{{\rm{Ref}}}{\bm{\omega }}_{{p_m}}^1 = {({\bm{\omega }}_{{q_m}}^1)^H}{\bf{N}}_k^m{\bf{\Theta }}_k^m{\bf{M}}_k^m{\bm{\omega }}_{{p_m}}^1,
\end{equation}
where $|{({\bf{\omega }}_{{q_m}}^1)^H}{\bf{N}}_k^m|{|_2} \gg 0$ and $||{\bf{M}}_k^m{\bf{\omega }}_{{p_m}}^1|{|_2} \gg 0$. During the interval when this beam pair (aligned case) is used, user$k$ will detect an energy pulse in the time slot when IRS$l_k^m$ uses ${\bf{\Theta }}_k^m={\bf{\Theta }}_{\rm{ref}}(\Delta _{k,m})$, i.e.,
\begin{equation}
\begin{split}
{\Delta _{k,m}} = &\arg \mathop {\max }\limits_x {({\bm{\omega }}_{{q_m}}^1)^H}{\bf{N}}_k^m{{\bf{\Theta }}_{{\rm{ref}}}}(x){{\bf{M}}_k^m}{\bm{\omega }}_{{p_m}}^1\\
&{\rm{s}}{\rm{.t}}{\rm{.}}\;\;\;x =  - 2 + \frac{{2i }}{N},\;\;\;i = 1,....,2N-1.
\end{split}
\end{equation}
We note that each IRS is able to use spatial modulation to label identity information\cite{tongji}. Thus, user$k$ can utilize the pulse slots plus identity information to get $\Delta _{k,m}$.
\subsubsection{Phase 2 (see Fig. \ref{figp2})}
In Phase 2, we turn off all IRSs and aim to obtain $\varphi _{B,H}^k$ and $\varphi _{U,H}^k$ via the following three steps. In step 1, $9$ wide-beam pairs in stage $1$ are tested for alignment and the resulting signals can be written in a vector form as
\begin{align}
{\bf{y}}_1^{{\rm{Ues}}}&=\sqrt P  {\rm{vec(}}{G_tG_r\bf{Q}}_1^H{\bf{H}}_k^{{\rm{LoS}}}{{\bf{P}}_1}+{{\bf{N}}_{{\rm{noise}}}})\notag\\
 &= c_k \left[ {{\bf{P}}_1^T{\bf{a}}_{{N_t}}^B{{\left( {\varphi _{B,H}^k} \right)}^\dag }} \right] \otimes \left[ {{\bf{Q}}_1^H{\bf{a}}_{{N_u}}^U\left( {\varphi _{U,H}^k} \right)} \right] + {\bf{n}}.\label{jiegou}
\end{align}
\begin{figure}[t]
\center
\includegraphics[width=3.4in]{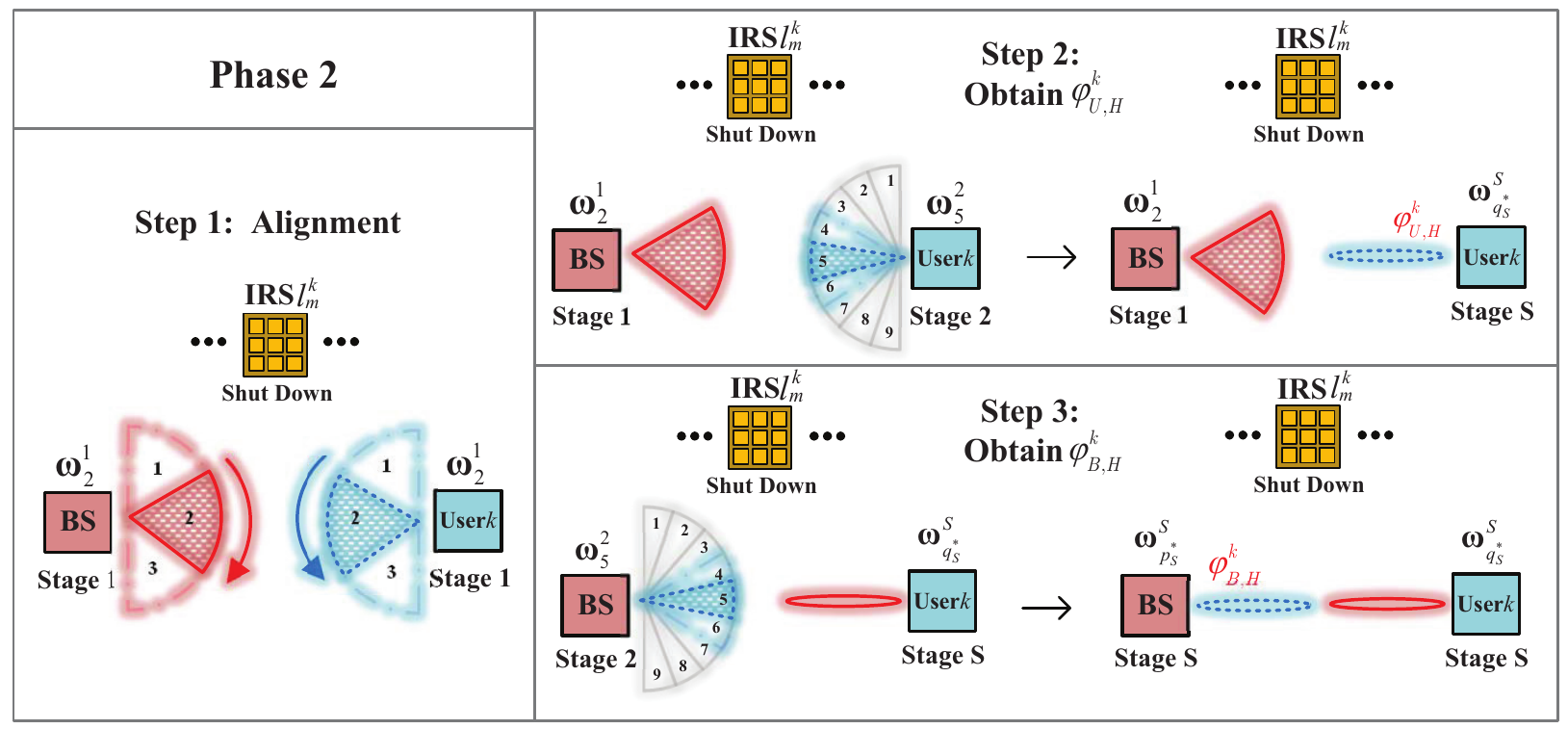}
\caption{Phase 2 of the cooperative beam training procedure.}\label{figp2}
\vspace{-12pt}
\end{figure}where ${{\bf{P}}_1} = {{\bf{Q}}_1} = [{\bm{\omega }}_1^1,{\bm{\omega }}_2^1,{\bm{\omega }}_3^1]$, ${{\bf{N}}_{{\rm{noise}}}}$ is the noise matrix and $c_k=\sqrt P G_tG_ra(f,{d_{k,0}})$. User$k$ compares the received energy in $9$ intervals and determines the aligned pair with the maximum power, i.e., $n^* = \arg {\max _n}[{\bf{y}}_1^{{\rm{Ues}}} \odot {({\bf{y}}_1^{{\rm{Ues}}})^\dag }](n)$. The aligned pair is labeled by recording the beam choice of both sides with $p_1^* = \left\lceil {n^*/3} \right\rceil $ and $q_1^* = n^*-3(p_1^*-1)$. In step 2, BS transmits the wide beam ${\bm{\omega }}_{p_1^*}^1$ and user$k$ uses the ternary-tree search by codewords in stage $s=2,...,S$ to obtain $\varphi _{U,H}^k$. Thus, the resulting signals in stage $s$ can be written as
\begin{equation}
{\bf{y}}_s^{{\rm{Ues}}} = \sqrt P {{G_t}{G_r}{\bf{Q}}_s^H{\bf{H}}_k^{{\rm{LoS}}}{\bm{\omega }}_{p_1^*}^1 + {{\bf{n}}}},
\end{equation}
where ${{\bf{Q}}_s} \!=\! [{\bm{\omega }}_{3q_{s \!-\! 1}^* \!-\! 2}^s, {\bm{\omega }}_{3q_{s \!-\! 1}^* \!-\! 1}^s,{\bm{\omega }}_{3q_{s \!-\! 1}^*}^s]$. Meanwhile, we calculate the parameters $q_s^* = 3q_{s - 1}^* + \arg {\max _n}[{\bf{y}}_s^{{\rm{Ues}}} \odot ({\bf{y}}_s^{{\rm{Ues}}})\dag ](n)$ to choose ${{\bf{Q}}_{s+1}}$ for the next stage. By recursively repeating the above procedure, user$k$ can find the desired narrow beam in stage $S$ and the angle $\varphi _{U,H}^k$ is estimated by the direction of ${\bm{\omega }}_{q_S^*}^S$. In step 3, 
user$k$ transmits the narrow beam ${\bm{\omega }}_{q_S^*}^S$ and BS uses the ternary-tree search in the same way to obtain $\varphi _{B,H}^k$ and the resulting signals in stage $s=2,...,S$ can be written as
\begin{equation}
{\bf{y}}_s^{{\rm{Bes}}} = \sqrt P {{G_t}{G_r}{\bf{P}}_s^H{{({\bf{H}}_k^{{\rm{LoS}}})}^T}{\bm{\omega }}_{q_S^*}^S + {{\bf{n}}}} ,
\end{equation}
where ${{\bf{P}}_s} \!=\! [{\bm{\omega }}_{3p_{s \!-\! 1}^* \!-\! 2}^s, {\bm{\omega }}_{3p_{s \!-\! 1}^* \!-\! 1}^s,{\bm{\omega }}_{3p_{s \!-\! 1}^*}^s]$ and we calculate the parameters $p_s^* = 3p_{s - 1}^* + \arg {\max _n}[{\bf{y}}_s^{{\rm{Bes}}} \odot ({\bf{y}}_s^{{\rm{Bes}}})\dag ](n)$ to choose ${{\bf{P}}_{s+1}}$ for the next stage. Likewise, the angle $\varphi _{U,H}^k$ is estimated by the direction of ${\bm{\omega }}_{p_S^*}^S$, and the path loss $a(f,{d_{k,0}})$ can be detected by the received energy via narrow-beam pair.
\begin{algorithm}
  \caption{Cooperative Beam Training Procedure for user$k$ in IRS-assisted THz MIMO Systems.}
 \textbf{Input:} Number of IRS $N_i$, number of narrow beams $N$; hierarchical codebook ${\rm{\{ }}{\bm{\omega }}_n^s{\rm{\} }}_{s = 1}^S$, $S = \log _3^N$.\\
\textbf{Phase 1:}\\
\For{$p = 1:3$}
{
  BS uses precoder ${\bm{\omega }}_p^1$.\\
  \For{$q = 1:3$}
  {
  User$k$ uses combiner ${\bm{\omega }}_q^1$\\
    \For{$i = 1:2N-1$}
     { 
      All IRSs turn to ${\bf{\Theta }}_{\rm{ref}}(- 2 + \frac{{2i}}{N})$.\\
      \If{\rm{user$k$ detects a pulse with identity $m$.}}
      {\textbf{Output result:}  ${\Delta _{k,m}} =  - 2 + \frac{{2i}}{N}$.}
     }
  }
}
\textbf{Phase 2:} Turn off all IRSs. \\  

/$*$ After 9 tests for alignment: $*$/
$ \qquad {\bf{y}}_s^{{\rm{Ues}}} = \sqrt P {\rm{vec(}}{\bf{Q}}_s^H{\bf{H}}_k^{}{{\bf{P}}_s} +{{\bf{N}}_{{\rm{noise}}}})$.\\
$\qquad n^* = \arg {\max _n}[{\bf{y}}_s^{{\rm{Ues}}} \odot {({\bf{y}}_s^{{\rm{Ues}}})^\dag }](n),$\\
$\qquad p_1^* = \left\lfloor {n^*/3} \right\rfloor $ and $q_1^* = n^*-3(p_1^*-1)$.\\
\For{\rm{stage} $s = 2:S$}
{
  BS uses precoder ${\bm{\omega }}_{p_1^*}^1$.\\
  \For{$q = {3q_{s \!-\! 1}^* \!-\! 2}:{3q_{s \!-\! 1}^*}$}
  {
  User$k$ uses combiner ${\bm{\omega }}_q^s$.
  }
${\bf{y}}_s^{{\rm{Ues}}} = \sqrt P {\bf{Q}}_s^H{\bf{H}}_k{\bm{\omega }}_{p_1^*}^1 + {{\bf{n}}}$.\\
$q_s^* = 3q_{s - 1}^* + \arg {\max _n}[{\bf{y}}_s^{{\rm{Ues}}} \odot ({\bf{y}}_s^{{\rm{Ues}}})\dag ](n)$.
}
\For{\rm{stage} $s = 2:S$}
{
  User$k$ uses precoder ${\bm{\omega }}_{q_S^*}^S$.\\
  \For{$p = {3p_{s \!-\! 1}^* \!-\! 2}:{3p_{s \!-\! 1}^*}$}
  {
  BS uses combiner ${\bm{\omega }}_p^s$.
  }
${\bf{y}}_s^{{\rm{Bes}}} = \sqrt P {\bf{P}}_s^H{\bf{H}}_k^T{\bm{\omega }}_{q_S^*}^S + {{\bf{n}}}$.\\
$p_s^* = 3p_{s - 1}^* + \arg {\max _n}[{\bf{y}}_s^{{\rm{Bes}}} \odot ({\bf{y}}_s^{{\rm{Bes}}})\dag ](n)$.
}
${\bf{m}}_p \!=\! {\bm{\omega }}_{p_S^*}^S,{\bf{m}}_q \!=\! {\bm{\omega }}_{q_S^*}^S$, calculate $a(f,{d_{k,0}})$.\\ 
\textbf{Phase 3:}  \\  
\For{$l = 1:N_i$}
{Turn on IRS$l_k^m$ and shut down other IRS. \\
/$*$ After 9 tests for alignment: $*$/\\
Calculate (\ref{ph31}).\\
$\qquad n^* = \arg {\max _n}[{\bf{y}}_s^{{\rm{Ues}}} \odot {({\bf{y}}_s^{{\rm{Ues}}})^\dag }](n),$\\
$\qquad p_1^* = \left\lfloor {n^*/3} \right\rfloor $ and $q_1^* = n^*-3(p_1^*-1)$.\\
\For{\rm{stage} $s = 2:S$}
{
  BS uses precoder ${\bm{\omega }}_{p_1^*}^1$.\\
  \For{$q = {3q_{s \!-\! 1}^* \!-\! 2}:{3q_{s \!-\! 1}^*}$}
  {
  User$k$ uses combiner ${\bm{\omega }}_q^s$.
  }
Calculate (\ref{ph32}).
$q_s^* = 3q_{s - 1}^* + \arg {\max _n}[{\bf{y}}_s^{{\rm{Ues}}} \odot ({\bf{y}}_s^{{\rm{Ues}}})\dag ](n)$.
}
\For{\rm{stage} $s = 2:S$}
{
  User$k$ uses precoder ${\bm{\omega }}_{q_S^*}^S$.\\
  \For{$p = {3p_{s \!-\! 1}^* \!-\! 2}:{3p_{s \!-\! 1}^*}$}
  {
  BS uses combiner ${\bm{\omega }}_p^s$.
  }
Calculate (\ref{ph33}).
$p_s^* = 3p_{s - 1}^* + \arg {\max _n}[{\bf{y}}_s^{{\rm{Bes}}} \odot ({\bf{y}}_s^{{\rm{Bes}}})\dag ](n)$.
}
Record the directions of $ {\bm{\omega }}_{p_S^*}^S$ and ${\bm{\omega }}_{q_S^*}^S$
}
\textbf{Output result:} Path angles and path loss. 
\end{algorithm}

\subsubsection{Phase 3 (see Fig. \ref{figp3})}
\begin{figure}[t]
\center
\includegraphics[width=3.4in]{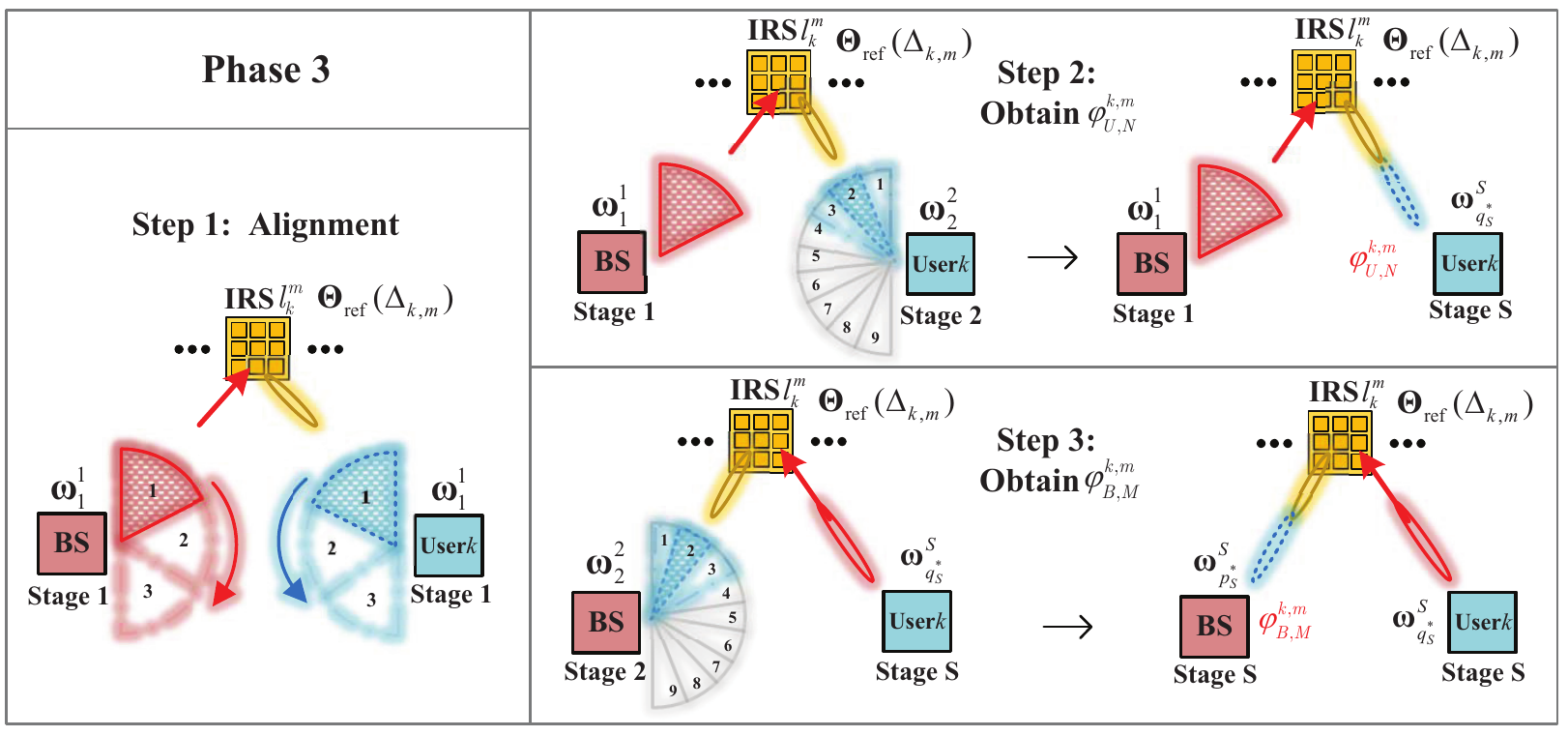}
\caption{Phase 3 of the cooperative beam training procedure.}\label{figp3}
\vspace{-12pt}
\end{figure}
In Phase 3, we aim to obtain $\varphi _{B,M}^{k,m}$ and $\varphi _{U,N}^{k,m}$ through three steps similar to Phase 2. With the obtained $\Delta _{k,m}$  in Phase 1, we successively turn on IRS with reflecting mode ${\bf{\Theta }}_k^m={\bf{\Theta }}_{\rm{ref}}(\Delta _{k,m})$ in $N_i$ intervals. In each interval, there exist two propagation paths from BS to user$k$, i.e., BS-IRS$l_k^m$-user$k$ path and BS-user$k$ path. Note that the BS-user$k$ path has been estimated in Phase 2. Hence, we assign the corresponding codewords (${\bm{\omega }}_{p_S^*}^S$ and ${\bm{\omega }}_{q_S^*}^S$)  to vectors ${\bf{m}}_p$ and ${\bf{m}}_q$, respectively. To estimate the AoA and AoD of the reflecting paths, we can use the ternary-tree search with small modification. Specifically, in each stage, the receiver removes the signal from the  BS-user$k$ path when determining the best beam. By this means, the effective received signals in step 1 are calculated as
\begin{align}\label{ph31}
{\bf{y}}_1^{{\rm{Ues}}} &= \underbrace {\sqrt P {\rm{vec(}}{\bf{Q}}_1^H{\bf{H}}_k^{}{{\bf{P}}_1}{\rm{ + }}{{\bf{N}}_{{\rm{noise}}}}{\text{)}}}_{{\text{received}}\;{\text{power}}}\\
&\;\;\;\;\;\;\;\;\; - \underbrace {{\sqrt P {\rm{vec}}\left( {{G_t}{G_r}a(f,{d_{k,0}}){\bf{Q}}_s^H{\bf{m}}_q{{({\bf{m}}_p)}^H}{{\bf{P}}_1}} \right)} }_{{\text{BS-user}}k\;{\text{signal}}}.\notag
\end{align}
In step 2, the effective received signals are calculated as
\begin{align}\label{ph32}
{\bf{y}}_s^{{\rm{Ues}}} &= \sqrt P {\bf{Q}}_s^H{\bf{H}}_k{\bm{\omega }}_{p_1^*}^1 + {{\bf{n}}}\\
&\;\;\;\;\;\;\;\;\; -  {\sqrt P {{G_t}{G_r}a(f,{d_{k,0}}){\bf{Q}}_s^H{\bf{m}}_q{{({\bf{m}}_p)}^H}{\bm{\omega }}_{p_1^*}^1}} .\notag
\end{align}
In step 3, the effective received signals are calculated as
\begin{align}\label{ph33}
{\bf{y}}_s^{{\rm{Bes}}} &= \sqrt P {\bf{P}}_s^H{\bf{H}}_k^T{\bm{\omega }}_{q_S^*}^S + {{\bf{n}}}\\
&\;\;\;\;\;\;\;\;\; - {\sqrt P {{G_t}{G_r}a(f,{d_{k,0}}){\bf{P}}_s^H({\bf{m}}_p)^{\dag}{{({\bf{m}}_q)}^T}{\bm{\omega }}_{q_S^*}^S} } .\notag
\end{align}
After $N_i$ intervals, we find $N_i$ narrow-beam pairs in the reflecting paths. The angles $\varphi _{B,M}^l$, $\varphi _{U,N}^{l,k}$ and path loss can be obtained by the corresponding narrow-beam pair. 

The core process of the whole channel estimation procedure for user$k$ in IRS-assisted THz MIMO systems is summarized in Algorithm 1. 
\subsection{Complexity Analysis}
\begin{table*}[t]
\centering
\caption{Comparison of beam training technologies.}
\vspace{-5pt}
\hspace{0.4cm}
\begin{tabular}{|c|c|c|c|c|}
\hline
Beam training approach & \tabincell{c}{Applicability to\\ IRS-assisted system}  & \tabincell{c}{Applicability to\\ THz communication}  & \tabincell{c}{Search times for a\\three-node IRS-assisted system} & \tabincell{c}{Search times for a\\point-to-point system} \\ \hline
Exhaustive search     & Yes  & Yes   & ${N^2} + {N^4}$  & ${N^2} $      \\ \hline
One-side search\cite{ones}      & No & No  & $-$ &  $2N$\\ \hline
Adaptive binary-tree search\cite{track2}  & No & No  &  $-$  & $4\log _2N$ \\ \hline
Parallel beam search\cite{parallel}& No  & Yes  &  $-$   & $\left. N^2 \middle/ N_{RF} \right.$ \\ \hline
Two-stage training scheme\cite{thzh}& No &  Yes & $-$  &  $\left. N^2  \middle/M \right. + M$ \\ \hline
Proposed procedure via TD& Yes &  Yes   &  $18N + 12\log _3N - 3$  &  $6\log _3N + 3$ \\ \hline
Proposed procedure via PSD& Yes  & Yes   &  $6N+ 4\log _3N - 1$   &  $2\log _3N + 1$ \\ \hline
\end{tabular}
\end{table*}
In this subsection, we give the search complexity of our proposed training procedure. For a basic three-node IRS-assisted system as shown in Fig. \ref{fig3n}, the search times of the exhaustive beam training are $N^2+N^4$, whereas that of our proposed procedure via TD codebook are $18N + 12\log _3N - 3$ ($18N-9$ in Phase 1 and $6\log _3N + 3$ in both Phases 2 and 3). Assume that the number of RF chains is greater than 3, i.e., $N_{RF}\ge 3$. In our proposed procedure via the PSD codebook, in each stage, the receiver can detect the incoming signals in three directions simultaneously by using three RF chains. As such, its search times are $6N + 4\log _3N - 1$ ($6N-3$ in Phase 1 and $2\log _3N + 1$ in both Phase 2 and 3). For example, when $N=27$, the exhaustive beam training needs $532170$ tests, but the proposed procedure via TD (resp. PSD) only needs $519$ (resp. $173$) tests, which is much more efficient. 

It is worth mentioning that our proposed training procedure can also be applied to the traditional point-to-point system without IRS by only implementing the steps in Phase 2. In this case, the search times of our proposed procedure via TD (resp. PSD) are reduced to $6\log _3N + 3$ (resp. $2\log _3N + 1$). Here, we compare the applicability of different beam training approaches to IRS-assisted system and THz communications, as well as their complexity in Table 1. Specifically, the beam training in IEEE
802.11ad utilizes an one-sided search algorithm \cite{ones}, where each user exhaustively searches the beams in the codebook while the BS transmits the signal in an omnidirectional mode. Adaptive binary-tree search realizes the beam training by selecting beam pairs stage by stage with decreasing beam width \emph{on both sides}
and the complexity is given by $4\log _2N$\cite{track2}. The authors in \cite{parallel} proposed a parallel beam search approach which uses $N_{RF}$
RF chains at BS to transmit multiple beams simultaneously while all users exhaustively search the beams, which incurs the complexity of $\left. N^2 \middle/ N_{RF} \right.$. Compared to the scheme in \cite{track2}, the parallel beam search in our procedure with multiple RF chains for detection has the following advantages. First, the transmit signal does not need to carry information to distinguish the transmitted direction for receivers. Second, after the alignment in step 1, our procedure does not need feedback between stages, while the scheme of transmitting multiple beams need feedback in every stage to initial the next stage. The authors in \cite{thzh} proposed a two-stage training scheme which combines sector level sweeping and hierarchical search for THz communication and results in the complexity of $\left. N^2  \middle/M \right. + M$, where $M$ is the number of narrow beams covered in each sector level. As noted from Table 1, except the exhaustive search, only our proposed approaches can be applied in THz IRS-assisted system. Besides, our proposed approaches are also efficient for traditional point-to-point communication system.

\section{Designs of IRSs and Hybrid Precoder/Combiners}\label{dede}
Given the beam training results,  in this section, we aim to design the IRS-PSs and hybrid precoder/combiners at the BS/users to maximize the achievable rate of the considered downlink IRS-assisted MIMO system.
\subsection{Design of IRS-PSs and Analog Precoder/Combiners}
We assume that all BS-user paths can be estimated (i.e., no obstacles between BS and users) and $N_{RF}^t=K(N_i+1)$. Let $\hat{\varphi} _{B,M}^{k,m}$, $\hat{\varphi} _{B,H}^k, \hat{\varphi} _{U,H}^k,$ and $\hat{\varphi} _{U,N}^{k,m}$ denote the estimated quantized angles for user$k$. Since the optimal IRS codewords and the beam codewords in (\ref{pran}) and (\ref{coman}) have been directly obtained via the beam training, the IRS-PS designs are given by
\begin{equation}\label{IRSd}
{{\bf{\Theta }}_k^m} = {{\bf{\Theta }}_{{\rm{ref}}}}({\Delta _{k,m}}),\;k=1,2,...,K,\;m=1,2,...,N_i.
\end{equation}
The BS's analog precoder and the users' analog combiners are given by
\begin{align}
&{{\bf{F}}_{RF}}= \bigg[{\bf{a}}_{{N_t}}^B\left( {\hat \varphi _{B,H}^1} \right),{\bf{a}}_{{N_t}}^B\left( {\hat \varphi _{B,M}^{1,1}} \right),...,{\bf{a}}_{{N_t}}^B\left( {\hat \varphi _{B,M}^{1,{N_i}}} \right),\notag\\
&\;\;\;\;\quad\qquad{\bf{a}}_{{N_t}}^B\left( {\hat \varphi _{B,H}^2} \right),{\bf{a}}_{{N_t}}^B\left( {\hat \varphi _{B,M}^{2,1}} \right),...,{\bf{a}}_{{N_t}}^B\left( {\hat \varphi _{B,M}^{2,{N_i}}} \right),\label{ds}\notag\\
&\;\;\;\qquad\qquad...,\\
&\;\qquad\qquad{\bf{a}}_{{N_t}}^B\left( {\hat \varphi _{B,H}^K} \right),{\bf{a}}_{{N_t}}^B\left( {\hat \varphi _{B,M}^{K,1}} \right),...,{\bf{a}}_{{N_t}}^B\left( {\hat \varphi _{B,M}^{K,{N_i}}}\right)\bigg],\notag
\end{align}
and
\begin{equation}
\begin{split}
{\bf{W}}_{RF}^k &= \bigg[{\bf{a}}_{{N_u}}^B\left( {\hat \varphi _{U,H}^k} \right),{\bf{a}}_{{N_u}}^U\left( {\hat \varphi _{U,N}^{{k,1}}} \right),{\bf{a}}_{{N_u}}^U\left( {\hat \varphi _{U,N}^{{k,2}}} \right),\\
&\quad\qquad\qquad\quad ...,{\bf{a}}_{{N_u}}^U\left( {\hat \varphi _{U,N}^{{k,{N_i}}}} \right)\bigg], k=1,2,...,K,
\end{split}
\end{equation}
respectively. In the case that some paths cannot be estimated by the proposed beam training (due to the LoS blockage or idle users), we remove the corresponding ARVs in ${{\bf{F}}_{RF}}$ and ${\bf{W}}_{RF}^k $ and use fewer RF chains.

\subsection{Design of Digital Precoder/Combiners}
Next, we design the digital precoder/combiners. In particular, we first propose a DPA scheme for the single-user scenario and then apply the BD approach\cite{BD} to the multi-user scenario.  First, a key lemma is introduced as below.\begin{lemma}[{[15]}]
For a ULA system with azimuth AoA or AoD drawn independently from a continuous distribution, the transmit and receive ARVs are orthogonal, i.e., ${\bf{a}}({\phi ^k}) \bot {\rm{span}}(\{ {\bf{a}}({\phi ^l})|\forall l \ne k\} )$, as the number of antenna elements, $N_a$, tends to infinity.\end{lemma}

Lemma 2 indicates that for massive MIMO channels, propagation paths with different AoD/AoA combinations possess a property of approximate space orthogonal on account of the large antenna number.
\subsubsection{DPA}
Considering that only user$k$ is scheduled to be served, the analog precoder ${{\bf{F}}_{RF}}$ applies the corresponding $D_k$ ARVs and $N_{RF}^t=N_{RF}^{u,k}=D_k$. According to Lemma 2, it is easy to verify that the ARVs in ${{\bf{F}}_{RF}}$ and ${\bf{W}}_{RF}^k$ can be approximately viewed as the singular vectors of ${\bf{H}}_k$, which are equivalent to the SVD precoder/combiner design for point-to-point MIMO channel. Based on this fact, the BS digital precoder can be directly designed as a diagonal power allocation matrix, with its diagonal elements denoting the power allocated to the $D_k$ spatial paths. Such a DPA can be mathematically expressed as
\begin{equation}\label{dw}
{{\bf{F}}_B} \!=\!{{\bf{F}}_B^k} \!=\! {\rm{diag}}\Big(\sqrt {{S_1}} ,\sqrt {{S_2}} ,...,\sqrt {{S_{D_k}}}\Big),\;{\bf{W}}_B^k \!=\! {{\bf{I}}_{D_k}},
\end{equation}
where $\{S_i\}_{i=1}^{D_k}$ are the power allocation factors. As such, the achievable rate of user$k$ reduces to
\begin{align}
{R_k} &=  {\log _2}{\rm{det}}\left[ {{{\bf{I}}_{{D_k}}} + \frac{P}{{\sigma _n^2}}{\bf{W}}_k^H{{\bf{H}}_k}{{\bf{F}}_k}{\bf{F}}_k^H{\bf{H}}_k^H{{\bf{W}}_k}} \right]\notag\\
 &= {\log _2}{\rm{det}}\left[ {{\bf{I}}_{{D_k}}} + \frac{P}{{\sigma _n^2}}({{\bf{F}}_B^k})^H{\rm{diag}}({{\bf{a}}_{{\rm{loss}}}}){{\bf{F}}_B^k}\right],
\end{align}
where ${{\bf{F}}_k}\!=\!{{\bf{F}}_{RF}}{{\bf{F}}_{B}^k}$ is the effective precoder component for user$k$ and ${{\bf{a}}_{{\rm{loss}}}}$ is the path loss vector. The remaining task is to determine the power allocation factor over these paths for maximizing the achievable rate, i.e.,
\begin{equation}\label{wf}
\begin{split}
&\mathop {\max }\limits_{{S_i}} \sum\limits_{i = 1}^{{D_k}} \log _2 \left[{1 + \frac{P}{\sigma _n^2}{{\bf{a}}_{\rm{loss}}(i)^2}{S_i}}\right] \\
&\;\;{\rm{s}}.{\rm{t}}.\;\;\;\;\sum\nolimits_{i = 1}^{{D_k}} {{S_i}}  = 1,\quad \{ {S_i}\} _{i = 1}^{{D_k}} \ge 0.
\end{split}
\end{equation}
Problem (\ref{wf}) is a classic water-filling problem and the optimal solution is given by
\begin{equation}\label{power}
{S_i} = \left\lceil {\frac{1}{{\ln 2 \cdot \mu }} - \frac{{\sigma _n^2}}{{P{\bf{a}}_{\rm{loss}}(i)^2}}} \right\rceil^+ ,\;\;\;i = 1,2,...,{D_k},
\end{equation}
where ${\left\lceil  \cdot  \right\rceil ^ + } = \max \{\cdot,0\}$. The parameter $\mu > 0$ is determined by the bisection search to satisfy the constraint $\sum\nolimits_{i = 1}^{{D_k}} {{S_i}}=1$.
\subsubsection{BD}
For the case that multiple users are served simultaneously, the achievable rate of user$k$ is given by \cite{rec2}
\begin{equation}\label{se}
{R_k} = {\log _2}{\rm{det}}\left[ {{{\bf{I}}_{{N_u}}} + P{{\bf{W}}_k}{\bf{C}}_k^{ - 1}{\bf{W}}_k^H{{\bf{H}}_k}{{\bf{F}}_k}{\bf{F}}_k^H{\bf{H}}_k^H} \right],
\end{equation}
where ${{\bf{C}}_k} \!= \! P{\bf{W}}_k^H{{\bf{H}}_k}\left( {\sum\nolimits_{i \ne k} {{{\bf{F}}_i}{\bf{F}}_i^H} } \right){\bf{H}}_k^H{{\bf{W}}_k} \!+\! {\sigma _n^2}{\bf{W}}_k^H{{\bf{W}}_k}$ is the covariance of the effective interference plus noise at user$k$. In this case, a straightforward strategy is to extend the DPA and directly allocate power over the all paths between the BS and users. However, with this strategy, inter-user interference cannot be completely eliminated since these spatial paths are only \emph{approximately orthogonal} due to the finite number of antenna elements. With the increase in the transmit power, the inter-user interference by DPA will greatly impair the sum-rate of all users, as will be shown in Section \ref{SR}.

To overcome the weakness of DPA at the high transmit power regime, we propose to apply the BD \cite{BD} that utilizes the digital domain of the BS to eliminate the inter-user interference. First, the effective channel between BS's digital precoder and user$k$'s digital combiner can be written as an $N_{RF}^{u,k}\times N_{RF}^t$ matrix with $N_{RF}^t=N_s$ and $N_{RF}^{u,k}=D_k$:
\begin{align}
{{\bf{H}}_{{\rm{eff}},k}} &= {G_t}{G_r}{({\bf{W}}_{RF}^k)^H}\times\\
&\Big\{\underbrace {{\bf{W}}_{RF}^k{\rm{diag}}({{\bf{a}}_{{\rm{loss}}}})[{\bf{F}}_{RF}(:,{E_{k - 1}} + 1:{E_k})]^H}_{{\text{estimated}}\;{\text{channel\;for}}\;{\text{user}}k}\Big\}{{\bf{F}}_{RF}},\notag
\end{align}
where ${E_k} = \sum\nolimits_{i = 1}^K {{D_k}}$ and ${\bf{F}}_{RF}(:,{E_{k \!-\! 1}} \!+\! 1:{E_k})$ represents the columns of ${\bf{F}}_{RF}$ from the $(E_{k \!-\! 1} \!+\! 1)$th column to the ${E_k}$th column. Then, we formulate the zero-interference space for each user (should lie in the null space of the effective channel between the BS and the other users). For user$k$, we concatenate the effective channels of other users as
\begin{equation}
{\overline {\bf{H}} _{{\rm{eff,}}k}} = {[{\bf{H}}_{{\rm{eff,1}}}^T,...,\;{\bf{H}}_{{\rm{eff,k - 1}}}^T,{\bf{H}}_{{\rm{eff,k + 1}}}^T,...,{\bf{H}}_{{\rm{eff,K}}}^T]^T}.
\end{equation}
By applying the SVD on the concatenated channels ${\overline {\bf{H}} _{{\rm{eff,}}k}} $, which is an $(N_s-D_k)\times N_s$ matrix, we have
\begin{equation}
{\overline {\bf{H}} _{{\rm{eff,}}k}} = {\overline {\bf{U}} _k}{\overline {\bf{\Sigma }} _k}{\left[ {\overline {\bf{V}} _k^{{\rm{eff}}},\overline {\bf{V}} _k^{{\rm{zero}}}} \right]^H}.
\end{equation}
Let $L_k$ represent the rank of ${\overline {\bf{H}} _{{\rm{eff,}}k}} $ with $L_k \leq N_s - D_k$. As such,  the null-space orthogonal basis of ${\overline {\bf{H}} _{{\rm{eff,}}k}} $ should consist of the last ($N_s-L_k$) right singular vectors, i.e., ${\overline {\bf{V}} _k^{{\rm{zero}}}}$. Thus, the projection of ${{\bf{H}}_{{\rm{eff}},k}}$ into the null space of ${\overline {\bf{H}} _{{\rm{eff,}}k}}$ can be written as ${{\bf{H}}_{{\rm{eff}},k}} \overline {\bf{V}} _k^{{\rm{zero}}}$. Then, we apply the SVD on this projected channel to maximize the achievable rate of user$k$ without causing inter-user interference, i.e.,
\begin{equation}
{\overline {\bf{H}} _{{\rm{eff,}}k}}\overline {\bf{V}} _k^{{\rm{zero}}} \!=\! [{\bf{U}}_k^{{\rm{eff}}},{\bf{U}}_k^{{\rm{zero}}}]\left[ {\begin{array}{*{20}{c}}
{{{\bf{\Sigma }}_k}}&{\bf{0}}\\
{\bf{0}}&{\bf{0}}
\end{array}} \right]{[{\bf{V}}_k^{{\rm{eff}}},{\bf{V}}_k^{{\rm{zero}}}]^H},
\end{equation}
where ${\bf{U}}_k^{{\rm{eff}}}$ (resp. ${\bf{V}}_k^{{\rm{eff}}}$) represents the first $D_k$ left (resp. right) singular vectors. ${{{\bf{\Sigma }}_k}}$ is a diagonal matrix that contains $D_k$ elements.  Thus, such a BD design can be written as
\begin{align}
&{{\bf{F}}_B} = \left[ {\overline {\bf{V}} _1^{{\rm{zero}}}{\bf{V}}_1^{{\rm{eff}}},\overline {\bf{V}} _2^{{\rm{zero}}}{\bf{V}}_2^{{\rm{eff}}},...,\overline {\bf{V}} _K^{{\rm{zero}}}{\bf{V}}_K^{{\rm{eff}}}} \right]\notag\\
 &\qquad \qquad \qquad\times {\rm{diag}}(\sqrt {{S_1}} ,\sqrt {{S_2}} ,...,\sqrt {{S_{{N_s}}}} ),\label{58}\\
&{\bf{W}}_B^k \!=\! {\bf{U}}_k^{{\rm{eff}}},\qquad k=1,2,...,K.\label{59}
\end{align}
Substituting (\ref{58}) and (\ref{59}) into (\ref{se}), it is easy to verify that ${\bf{C}}_k=\sigma_n^2{\bf{I}}_{D_k}$. Thus, the sum rate, i.e., $\mathop {\max }\nolimits_{{S_i}} \mathop \sum \nolimits_{k = 1}^K {R_k}$, can be rewritten in a convex form similar to (\ref{wf}) and the allocated powers ($S_1,...,S_{N_s}$) at BS can be obtained via the water-filling power allocation over all the diagonal elements in $\{{{{\bf{\Sigma }}_k}}\}_{k=1}^K$ which is given by 
\begin{equation}
{S_i} = \left\lceil {\frac{1}{{\ln 2 \cdot \mu }} - \frac{{\sigma _n^2}}{{P\Sigma_i^2}}} \right\rceil^+ ,\;\;\;i = 1,2,...,{N_s},
\end{equation}
where $\{\Sigma_i\}_{i=1}^{N_s}$ are the $N_s$ diagonal elements of ${\rm{diag}}({{{\bf{\Sigma }}_1}},{{{\bf{\Sigma }}_2}},...,{{{\bf{\Sigma }}_K}})$ in order and $\mu \ge 0$ is determined by the bisection search to satisfy the constraint $\sum\nolimits_{i = 1}^{{N_s}} {{S_i}}=1$.
\begin{remark}
The proposed DPA and BD designs are dedicated to narrow-band beamforming. Despite this limitation, we will show in Section \ref{SR} that the proposed schemes can yield high spectral efficiency in THz communication due to both the massive MIMO and IRS gains. For wideband communications, the beam squint effect should be taken into consideration which is worth an in-depth investigation in the future work.
\end{remark}

\section{Simulation Results}\label{SR}
In this section, we provide the simulation results to illustrate the performance of the proposed beam training scheme with hierarchical codebooks and the proposed HB designs with IRS.

\subsection{Performance of the Proposed Beam Training}\label{CDB}
In our proposed beam training scheme, the TD codebook and PSD codebook share the same bottom-stage narrow beams (see (\ref{phi})) but differ in upper-stage wide beams. In this subsection, we study the beam patterns of the proposed TD codebook and PSD codebook and evaluate their performance by testing the correct detection rate (which means the probability of successful alignment in all stages), where the codebook in\cite{track2} is presented as a benchmark.  Unless otherwise specified, we set the number of antenna elements as $N_a=32$ and the number of narrow beams as $N=81$.

%
\begin{figure}[t]
\center
\includegraphics[width=3.3in]{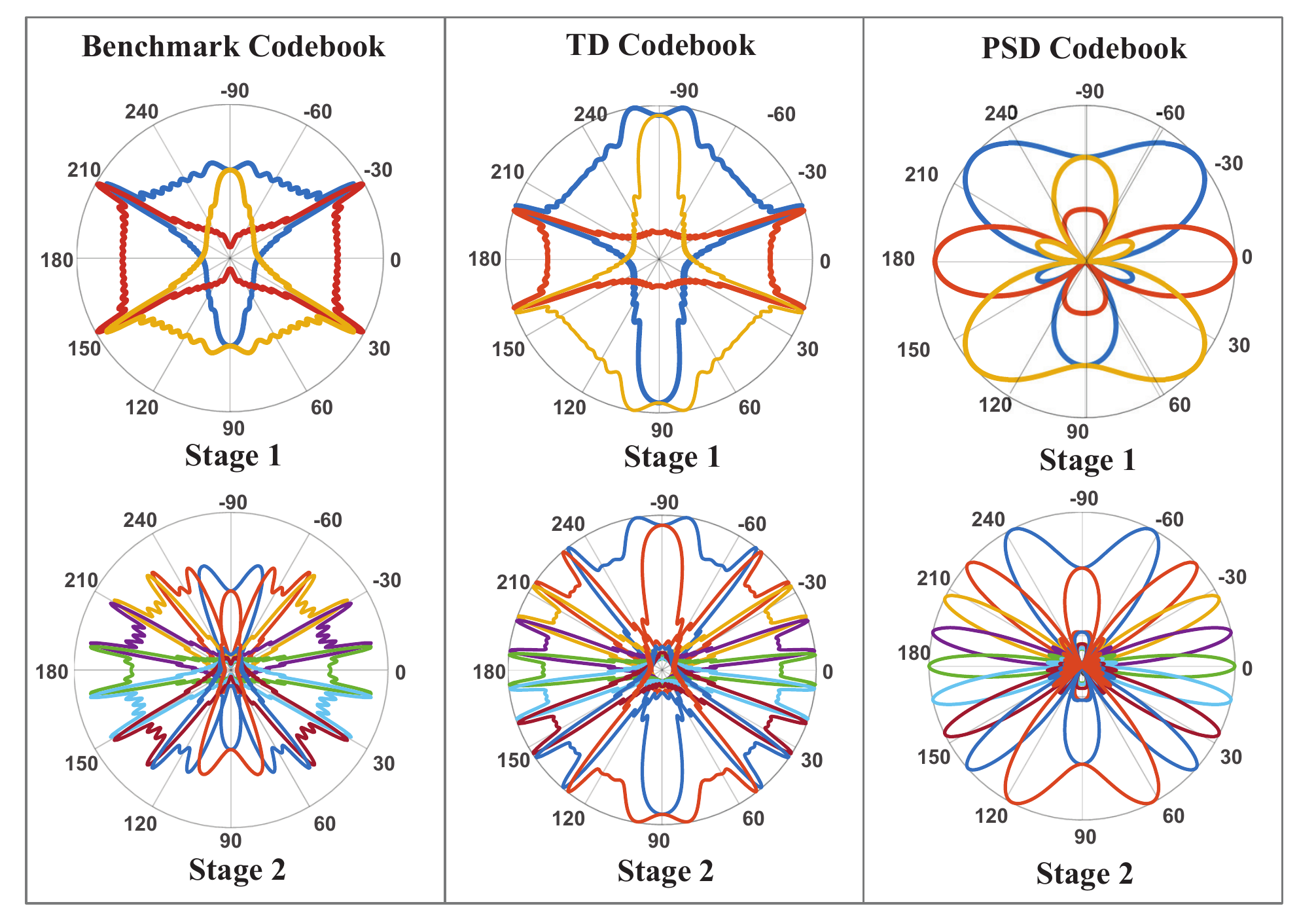}
\caption{Beam patterns of the proposed codebooks and that of the benchmark codebook\cite{track2}, where $N_a=32$ and $N=81$.}\label{compa}
\vspace{-5pt}
\end{figure}

Fig. \ref{compa} illustrates the beam patterns of different codebooks. The lines in different colors represent the beam patterns of different codewords in the same stage. For beams generated by the benchmark codebook, we observed that the beamwidth is identical for all beams on each side of array, e.g., $[-\pi /2,\pi /2]$, but the coverage-edge gain of each beam is different. On the contrary, the beams generated by the proposed codebooks have identical coverage-edge gain but different beamwidth. Comparing TD to PSD, we observe that the TD beams can better suppress the energy in the unintended directions. Thus, while the PSD codebook has the advantage of simultaneously testing multiple beams (since the realization of each codeword only needs one RF chain), the TD codebook is expected to show better anti-noise performance. \begin{figure}[t]
\center
\includegraphics[width=3.5in]{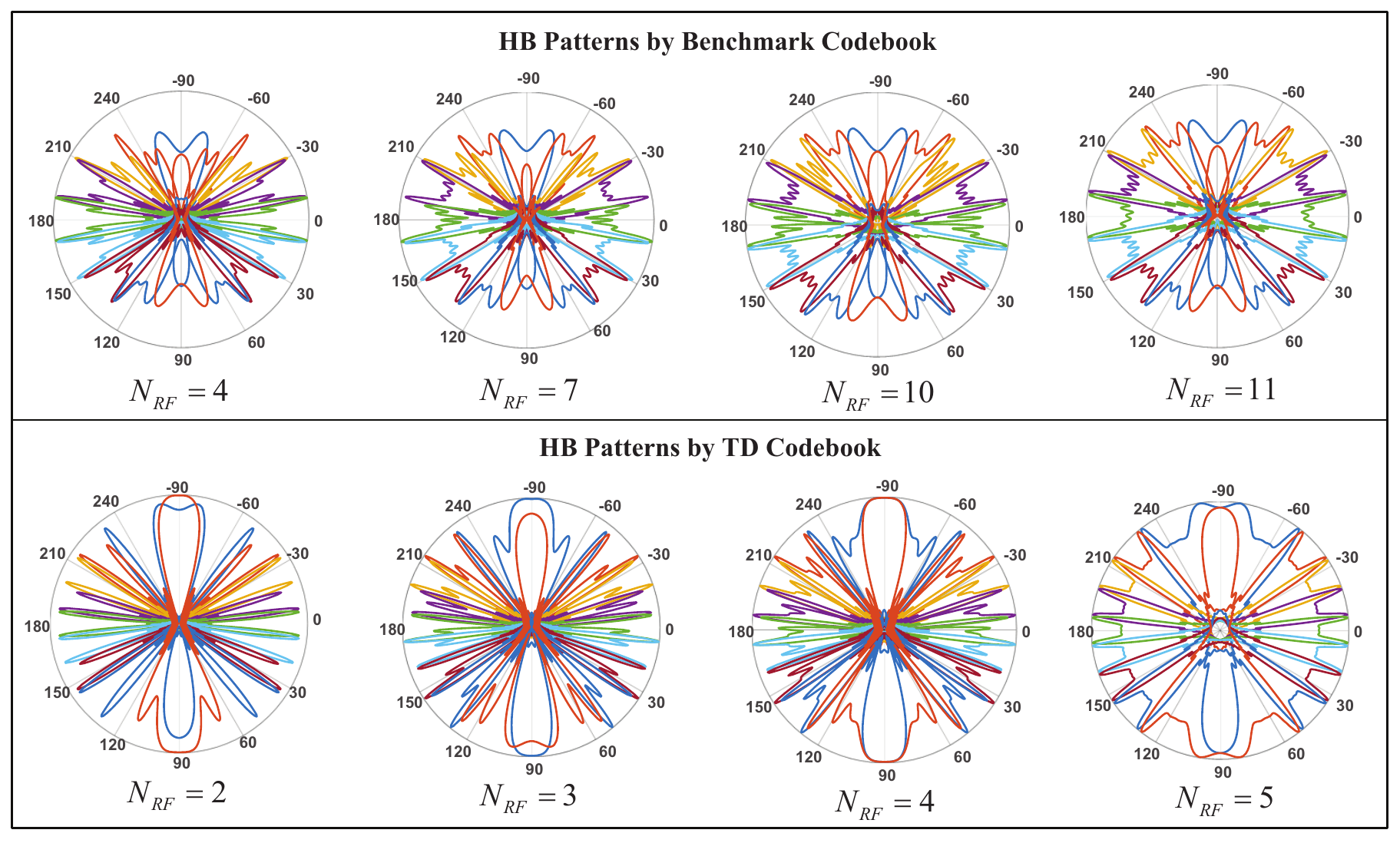}
\caption{HB patterns of the codewords in the second stage by using different numbers of RF chains, where $N_a=32$ and $N=81$.}\label{HBP}
\vspace{-12pt}
\end{figure}

It is worth mentioning that the HB with the benchmark codebook and the TD codebook are obtained by solving (\ref{omp}). To find the minimum number of RF chains needed to approach the codebook patterns shown in Fig. \ref{compa}, we test the HB patterns by adding RF chains successively in the simulation, and some of the test examples are demonstrated in Fig. \ref{HBP}. It is observed that the HB requires 11 RF chains to be close to the benchmark pattern in the second stage, whereas the HB requires only 5 RF chains to be close to the TD patterns. This reveals that the proposed TD codebook is cost-efficient for HB realization.
\begin{figure}[t]
\center
\includegraphics[width=3.2in]{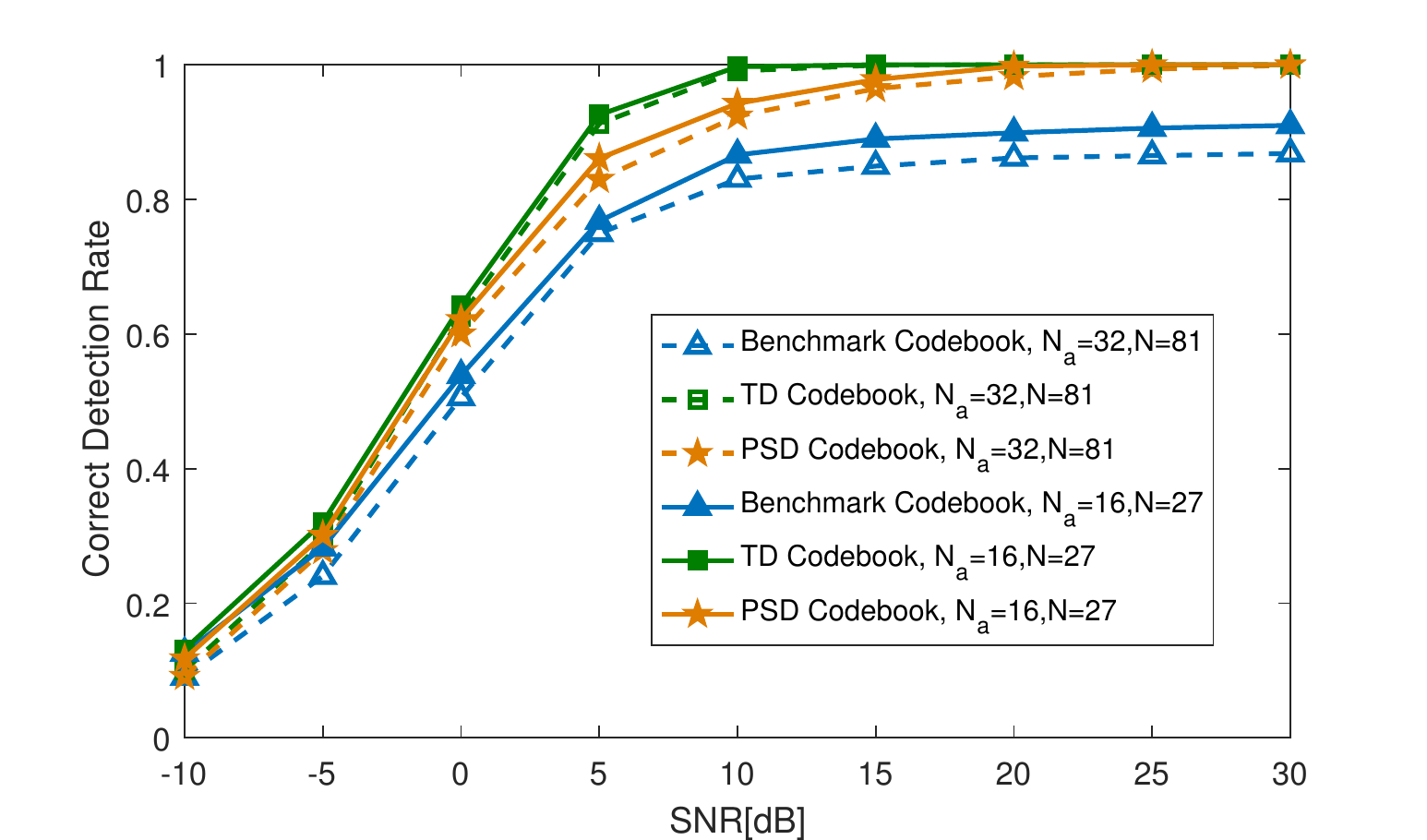}
\caption{The correct detection rate of using different codebooks. }\label{dtrate}
\vspace{-12pt}
\end{figure}

Next, we generate 100,000 training signals with different AoAs (i.e., ${{\bf{a}}_{{N_a}}}({\varphi})$ with different $\varphi$). For each signal ${{\bf{a}}_{{N_a}}}({\varphi})$, we label the desired narrow beam which covers $\varphi$. Then, we use the benchmark codebook as well as our proposed  codebooks to detect the directions of the training signals stage by stage respectively, in the presence of noises with different power. If the narrow beam found in the last stage exactly matches the desired one, then we call it a correct detection\cite{track3}. Let $N_{c}$ denote the number of correct detections. Then, the correct detection rate is given by $N_c/10^5$. Fig. {\ref{dtrate}} shows two sets of results (denoted by solid and dotted lines, respectively) of the correct detection rate by using different codebooks. Each set contains a performance of the three codebooks under the same training overhead and estimation accuracy\footnote{All codebooks are generated following the ternary-tree structure, thus $N$ represents the overhead and $N/N_a$ represents the estimation accuracy.}. It can be observed that our proposed codebooks can outperform the benchmark codebook in different setups. When the SNR is 30 dB, the proposed ternary-tree search with TD codebook and PSD codebook can both achieve $100\%$ correct detection.
Comparing the performance by the TD codebook and PSD codebook, we observed that by using TD codebook, the minimum SNR needed to achieve $100\%$ correct detection is lower than that of PSD, which validates the better anti-noise performance of the TD codebook.

\subsection{Performance of the IRS-assisted HB Schemes}
\begin{figure}[t]
\center
\includegraphics[width=3.2in]{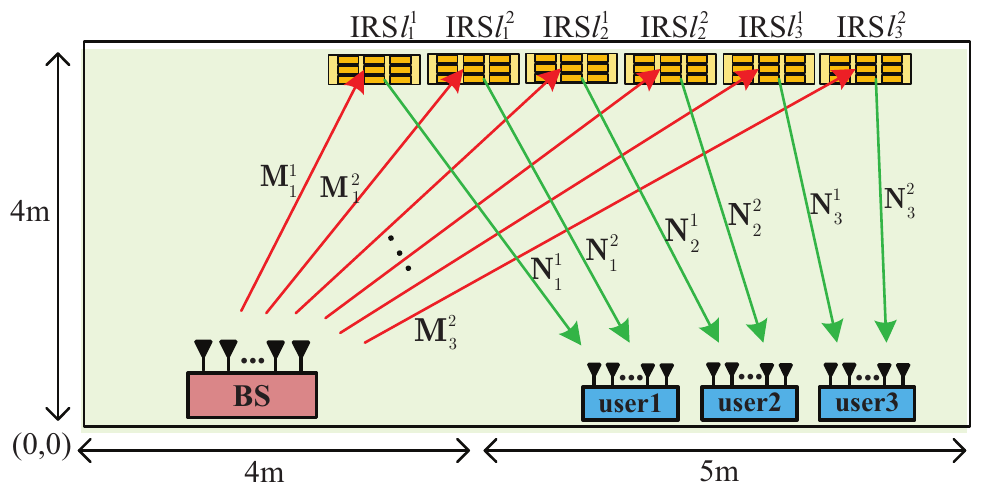}
\caption{The simulated IRS-assisted MIMO communication scenario.}\label{ind}
\vspace{-12pt}
\end{figure}
In this subsection, we evaluate the performance of the proposed HB schemes. As illustrated in Fig. \ref{ind},  we consider an indoor scenario where the BS is randomly located at $(0,{d_B})$ with $d_B \in [0,4]$ and $K=3$ users are randomly located at $(0,{d_{u_k}} \in [4,10])$. The $i$th IRS is located at $(4,2+i)$, $i=1,2,...,6$. The antenna spacing for the BS and all users is $d_a=\lambda/2$. The reflection coefficient and the compensation factor are set as $\beta=0.8$ and $\eta=1$ respectively. We assume the number of BS/user antennas and IRS elements are all identical $N_t\!=\!N_i\!=\!N_u$ , denoted as $N_a$. The operating frequency is set to $0.14$ THz and the background noise power at the receiver is $\sigma _n^2=-85$ dBm. Absorption coefficient is $\tau (0.14T)=1.83\times10^{-5}/m$ and antenna gains are considered as ${G_t}\!=\!{G_r} \!=\! 4+10\log _{10}(\sqrt{N_a})$. All the results presented were averaged over 10,000 random channel realizations. In particular, we plot the following schemes for comparison.
\begin{itemize}
\item \emph{FDB with optimal IRS-PSs}: It applies the optimal IRS-PSs and the fully digital beamforming (SVD for single-user scenario, zero-forcing precoding, for multi-user scenario).
\item \emph{Proposed HB with DPA}: It applies the proposed IRS-assisted HB with DPA digital precoder (for both single-user and multi-user scenario).
\item \emph{Proposed HB with BD}: It applies the proposed IRS-assisted HB with  BD digital precoder (only for multi-user scenario).
\item \emph{FDB with random IRS-PSs}: We randomly set the IRS-PSs and then apply fully digital beamforming (SVD for single-user scenario, zero-forcing precoding for multi-user scenario).
\item \emph{FDB without IRS}: In the scenario, we treat IRS as an indoor wall whose the first-order ray attenuations are between $5.8$ dB and $19.3$ dB compared to the LoS\cite{model}. Then, we perform fully digital beamforming (SVD for single-user scenario, zero-forcing precoding for multi-user scenario).
\end{itemize}
\subsubsection{Single-user Scenario}
\begin{figure}[t]
\center
\includegraphics[width=3.3in]{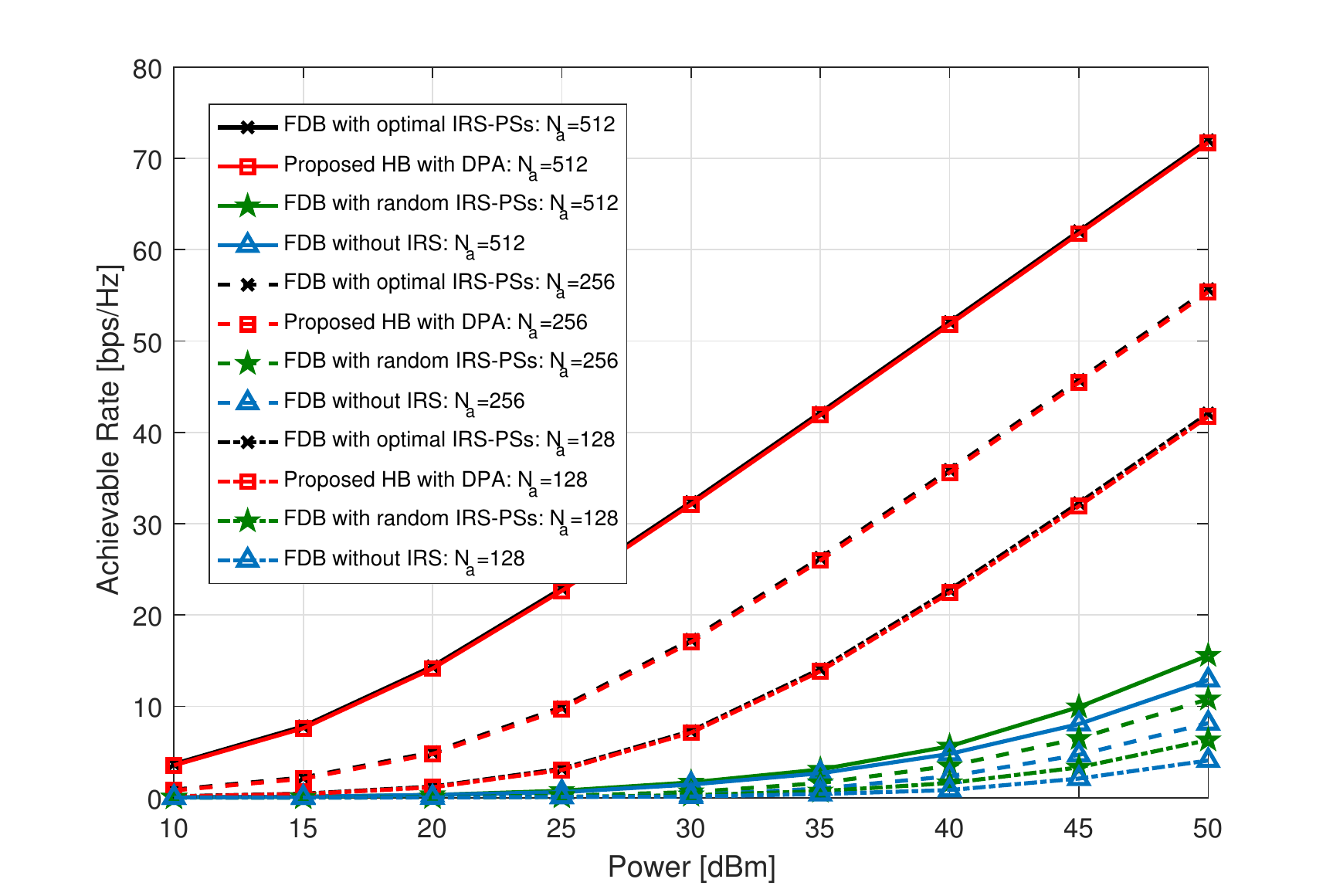}
\caption{Achievable rate versus transmit power in single-user scenario.}\label{n32}
 \vspace{-12pt}
\end{figure}
We first evaluate the IRS-assisted HB with DPA in a single-user scenario, where all six IRSs serve one user. Fig. \ref{n32} plots the achievable rate versus the transmit power for different schemes with $N_a=128$, $N_a=256$, and $N_a=512$. It is observed that the performance gains of the IRS-assisted schemes are significant compared to the schemes with random IRS-PSs and without IRS. This is expected as the IRSs are able to provide aperture gains via their controllable reflection, so as to increase the received power at the user. Besides, as the number of antennas $N_a$ increases from $128$ to $512$, the achievable rate by the proposed HB with DPA yields a significant improvement, which validates the applicability of the proposed scheme in massive MIMO systems.

\subsubsection{Multi-user Scenario}
\begin{figure}[t]
\center
\includegraphics[width=3.3in]{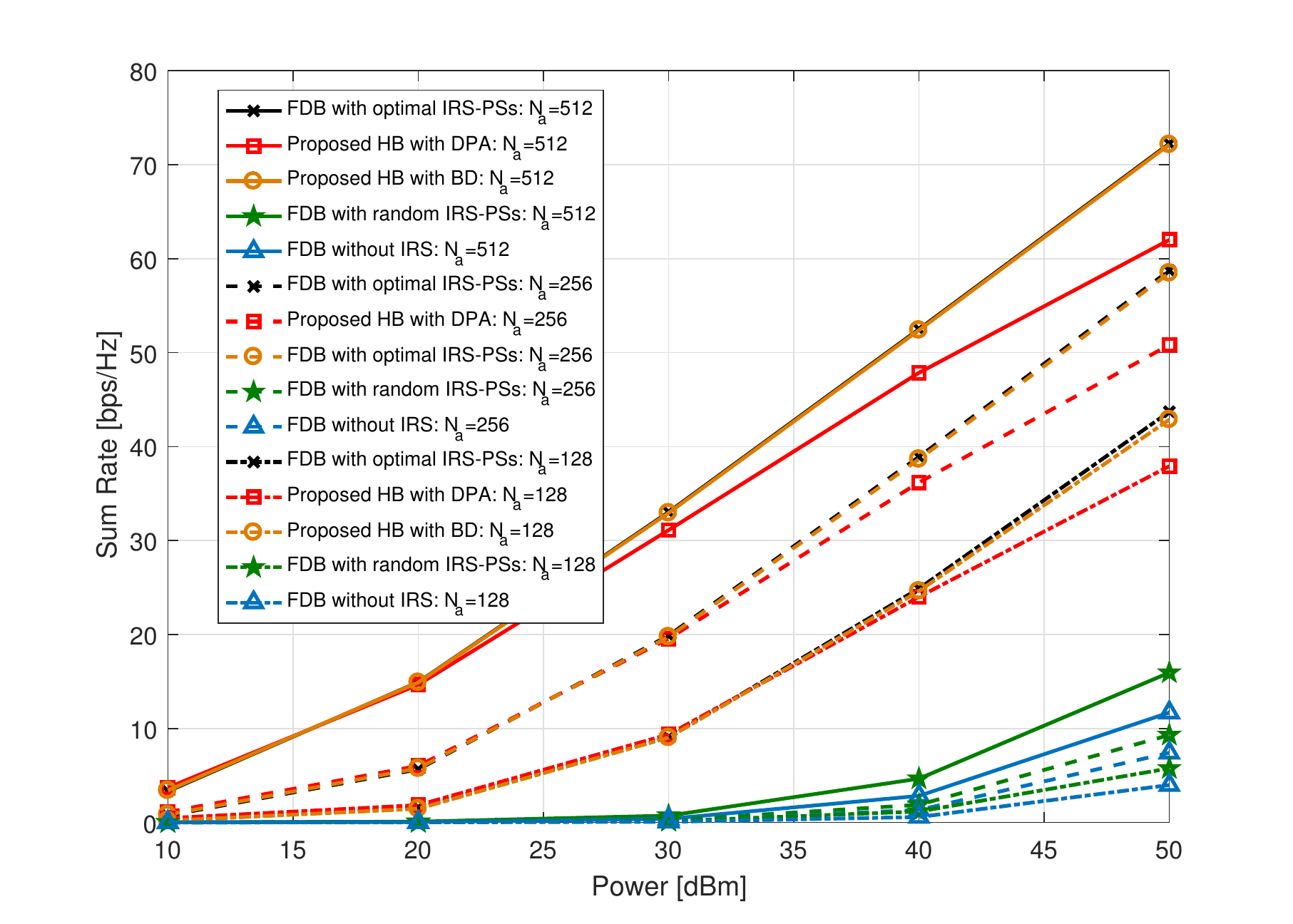}
\caption{Sum-rate versus transmit power in multi-user scenario.}\label{n64}
\vspace{-12pt}
\end{figure}

Next, we evaluate the IRS-assisted HB schemes in the multi-user scenario, where each user is served by two IRSs. Fig. \ref{n64} plots the sum-rate versus transmit power for different schemes with $N_a=128$, $N_a=256$, and $N_a=512$. It can be observed that the performance of HB with DPA is degraded at the high transmit power regime. This is because the inter-user interference, which cannot be eliminated in DPA, greatly compromises the sum rate. Compared to the DPA, the performance of HB with BD is closer to that of the fully digital beamforming baseline, which validates the effectiveness of BD in multi-user scenario. Moreover, Fig. \ref{n64} shows that the gap between HB with BD and fully digital beamforming baseline decreases with $N_a$. This can be explained as follows. The increase in $N_a$ leads to higher beam orthogonality which reduces the interference in the analog domain. As a result, the larger $N_a$ leads to larger null space for each user, and thus the BD performance approaches the zero-forcing baseline with increase of $N_a$. 
\begin{figure}[t]
\center
\includegraphics[width=3.0in]{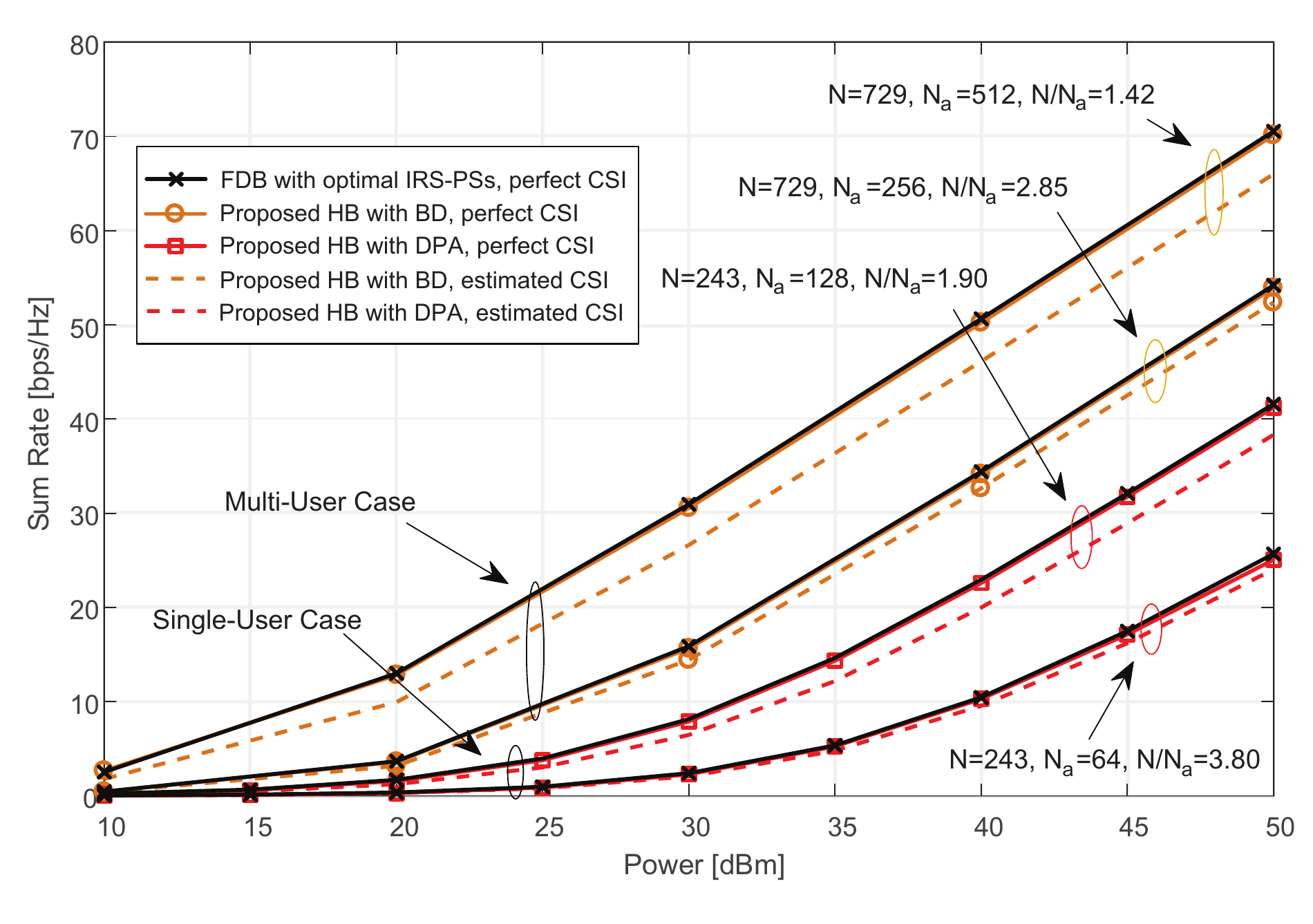}
\caption{Sum-rate versus transmit power for joint beam training and IRS-assisted HB schemes with different setups.}\label{n66}
\vspace{-12pt}
\end{figure} 
\begin{figure}[t]
\center
\includegraphics[width=3.3in]{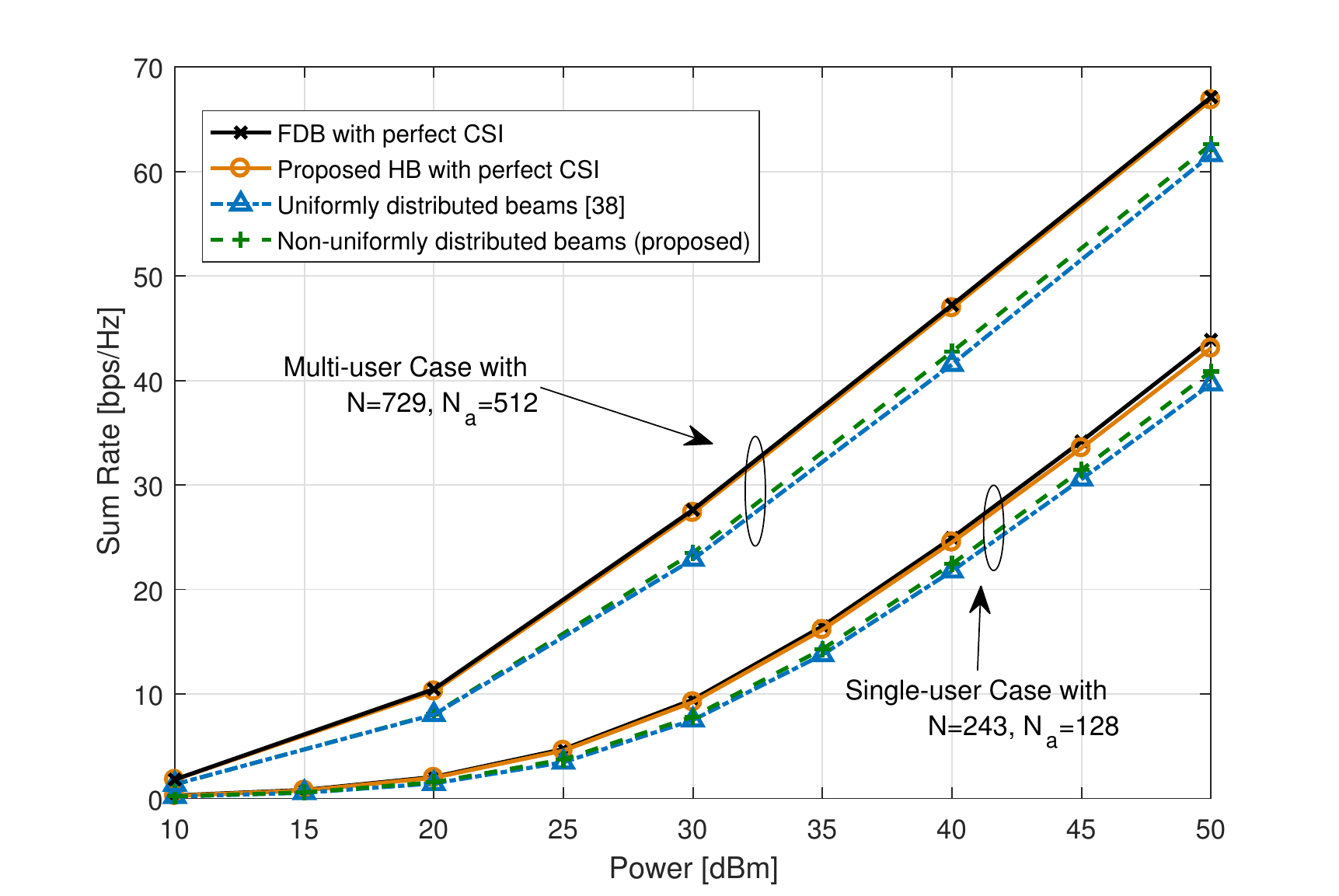}
\caption{Comparison of the sum-rate performance between the uniformly and the non-uniformly distributed beams.}\label{n67}
\vspace{-12pt}
\end{figure} 
\subsubsection{Joint Beam Training and IRS-assisted  HB}
Last, we evaluate the performance of IRS-assisted HB with DPA and BD with the estimated CSI obtained by the proposed beam training with $N$ narrow-beam candidates. We adopt a 5-stage and a 6-stage codebook, i.e., $N=3^5=243$ and $N=3^6=729$, in single-user and multi-user scenarios respectively. As seen from Fig. \ref{n66}, the performance is improved with the increase in antenna number $N_a$. However, with the fixed $N$, the larger $N_a$ brings a larger gap between the performance under perfect CSI and that under estimated CSI. This is because the performance loss caused by the estimated CSI is attributed to the AoDs/AoAs quantization error. Thus, the higher training accuracy $N/N_a$ indicates the less quantization error, and the less quantization error leads to a smaller gap to the performance under perfect CSI.
In Fig. \ref{n66}, there are four sets of lines, of which the training accuracy is $N/N_a=1.42$, $N/N_a=2.85$, $N/N_a=1.90$, and $N/N_a=3.80$, respectively. Obviously, the gap is the largest with $N/N_a=1.42$ and is the smallest with $N/N_a=3.80$. Finally, we compare the performance of our proposed joint beam training and HB beamforming scheme to that of the scheme in \cite{track2}. The training procedure of our proposed scheme is based on the non-uniformly distributed (or with the same coverage-edge) beams in (\ref{direc}), while the scheme in \cite{track2} is based on the uniformly distributed beams (or beams with the same beamwidth) in (\ref{nonui}). As can be seen in Fig. \ref{n67}, the performance of using the non-uniformly distributed beams outperforms that of using the uniformly distributed beams in both single-user and multi-user cases. It is also observed that the performance of the proposed joint beam training and HB design is close to that of the HB design with the perfect CSI, which validates the effectiveness of the proposed schemes.

\section{Conclusion}
We considered a low-complexity beam training and HB design for THz multi-user massive MIMO system with IRS. First, we established that there is no need to search the exact path AoD/AoA at IRS but only need to search the angle differences in the sine space. To further reduce the search complexity, an efficient ternary-tree search is proposed for the BS and users,  and it is proved to be more efficient compared to the binary-tree search. Further, we designed two novel codebooks, viz., TD codebook and PSD codebook, for implementing the ternary-tree search. In particular, the bottom-stage beams are distributed with the same coverage-edge gain, which is shown to lead to lower misalignment probability compared to the uniformly distributed beams. With the training results, we then proposed two cost-efficient HB designs with DPA and BD for maximizing the achievable rate. Numerical results showed that incorporating IRS yields significant performance improvement over the non-IRS-assisted counterparts and the proposed joint beam training and HB scheme are comparable to the fully digital beamforming scheme implemented under perfect CSI. Thus, with the implementation of the large-scale antenna array and adequate quantization resolution, the proposed scheme would be rather appealing in future THz massive MIMO network, due to its close-to-optimal performance and low complexity of hardware and signal processing.

\begin{appendices}    
\section{Proof of Proposition 1 and Corollary 1}
For narrow beam in direction $\varphi$, the normalized detective beam gain in (\ref{bpwer}) can be further expressed as
\begin{align}
A\left( {{{\bf{a}}_{{N_a}}}(\varphi ),\psi } \right) &= \left| {\frac{1}{{{N_a}}}\sum\limits_{n = 1}^{{N_a}} {{e^{jk{d_a}(n - 1)[\sin (\psi ) - \sin (\varphi )]}}} } \right|\notag\\
& =  \left| {\frac{1}{{{N_a}}}\frac{{{e^{j\frac{{{N_a}k{d_a}m}}{2}}}\left( {{e^{j\frac{{{N_a}k{d_a}m}}{2}}} - {e^{ - j\frac{{{N_a}k{d_a}m}}{2}}}} \right)}}{{{e^{j\frac{{k{d_a}m}}{2}}}\left( {{e^{j\frac{{k{d_a}m}}{2}}} - {e^{ - j\frac{{k{d_a}m}}{2}}}} \right)}}} \right|\notag\\
 &= \left| {\frac{1}{{{N_a}}}{e^{j\frac{{({N_a} - 1)k{d_a}m}}{2}}}\frac{{\sin [({N_a}k{d_a}m)/2]}}{{\sin [(k{d_a}m)/2]}}} \right|,
\end{align}
where $m = \sin (\psi ) - \sin (\varphi )$. Thus, for half-wavelength antenna spacing, i.e., $d_a=\lambda/2$, the normalized detective beam gain of ${{\bf{a}}_{{N_a}}}(\varphi _ n)$ is given by
\begin{equation}
A\left( {{{\bf{a}}_{{N_a}}}(\varphi _n ),\psi } \right) = \left| {\frac{{\sin [\frac{{{N_a}\pi }}{2}(\sin (\psi ) - \sin (\varphi _n))]}}{{{N_a}\sin [\frac{\pi }{2}(\sin (\psi ) - \sin (\varphi _n))]}}} \right|.
\end{equation}
Define a function ${\rm T}(x) = |\sin (\frac{{{N_a}\pi }}{2}x)/[ {N_a}\sin (\frac{\pi }{2}x)]|$. As shown in Fig. \ref{figsin}, it can be observed that the beam power is monotonically decreasing with the increase in $|x|$ when $|x| \leq 2/N_a$. Thus, the same $\rho$ yields an identical $|x|$ for all beams, and for any beam $n$, we have 
\begin{equation}
|\sin \varphi _n^{n + 1} - \sin {\varphi _n}| = |\sin {\varphi _n} - \sin \varphi _n^{n - 1}| = \chi(\rho),
\end{equation}
where $\chi$ is a constant and satisfy the inverse function $\rho={\rm T}(\chi)$. As $N$ beams cover all directions with the same $\rho$, we have $2N$ constant $\chi$ which uniformly divide the interval$[-1,1]$ with $\chi=1/N$ and the condition $N \ge {N_a}$ implies $\chi  <  1/{N_a}$. Thus, $\rho$ is monotonically increasing with the increase in $N$. In Fig. \ref{figpat}, $\sin {\varphi _1}\;,...,\sin {\varphi _{N}}$ uniformly divides the interval $[-1,1]$ and it is easy to obtain the beam direction by combining the feasible domain of the arcsine function.

Before proving Corollary 1, we first provide a key lemma which serves as the basis of the derivations that follows.
\begin{lemma}
Given two angle directions ${\varphi _1}$ and ${\varphi _2}$, we have 
\begin{equation}
\begin{split}
|\sin {\varphi _1} - \sin {\varphi _2}| &= \left| {2\cos \frac{{{\varphi _1} + {\varphi _2}}}{2}\sin \frac{{{\varphi _1} - {\varphi _2}}}{2}} \right|\\
&\mathop  \le 2\left| {\sin \frac{{{\varphi _1} - {\varphi _2}}}{2}} \right|\mathop  < \limits^{(a)} |{\varphi _1} - {\varphi _2}|,
\end{split}
\end{equation}
where $(a)$ is due to $|\sin x| < |x|$ if $|x|>0$.
\end{lemma}

For beams with the same beamwidth, the $i$th beam direction and the coverage-edge direction in FR is given by 
\[{\varphi _n} \!=\! \frac{({2n \!-\! 1 \!-\! N})\pi}{{2N}},\;\varphi _n^{n \!-\! 1} \!=\! \frac{{(2n \!-\! 2 \!-\! N)\pi }}{{2N}},\;\varphi _n^{n \!+\! 1} \!=\! \frac{{(2n \!-\! N)\pi }}{{2N}},\]
Then, by leveraging lemma 3, it follows that
\begin{align}
\rho  &= \min \left\{ {{\rm T}\left( {|\sin \varphi _n^{n + 1} - \sin {\varphi _n}|} \right),{\rm T}\left( {|\sin {\varphi _n} - \sin \varphi _n^{n - 1}|} \right)} \right\}\notag\\
 &> {\rm T}(\frac{\pi }{2N}) = {\rm T}(\frac{{\Delta \varphi }}{2}) = \frac{{\sin [(\Delta \varphi {N_a}\pi )/4]}}{{{N_a}\sin [(\Delta \varphi \pi )/4]}}
\end{align}
holds true for all beams\footnote{This is because ${\rm T}(x)$ is monotonically decreasing with $|x|$ when $|x| \leq 2/N_a$, and $N \ge {N_a}$ leads to $|\Delta \varphi /2| < 2/{N_a}$.}, which completes the proof.

\section{Proof of Proposition 4}
According to (\ref{c3}), the active antenna number for each wide beam is $N_{\rm{act}}=3^s$ for $s=1,2,...s_{\rm{max}}$. Let $ \varphi _n$ and $\varphi _n^e$ denote the beam direction and coverage-edge direction of ${\bm{\omega }}_n^s$. By setting $\rho$ of each beam in stage $s$ as in (\ref{p4}), we have 
\begin{align}
&\left| {\frac{{\sin \left[ {\frac{{{N_{{\rm{act}}}}\pi }}{2}\left| {\sin \varphi _n - \sin \varphi _n^e} \right|} \right]}}{{{N_{{\rm{act}}}}\sin \left[ {\frac{\pi }{2}\left| {\sin \varphi _n - \sin \varphi _n^e} \right|} \right]}}} \right|\;\\
&\qquad\qquad = \left| {\frac{{\sin \left[ {\frac{{{N_{{\rm{act}}}}\pi }}{2}\left( {\sin \varphi _n - \sin \varphi _n^e} \right)} \right]}}{{{N_{{\rm{act}}}}\sin \left[ {\frac{\pi }{2}\left( {\sin \varphi _n - \sin \varphi _n^e} \right)} \right]}}} \right|\notag\\
&\qquad\qquad = A({{\bf{a}}_{{N_{{\rm{act}}}}}}\left( {\varphi _n} \right),\varphi _n^e) = \rho (s) = \frac{1}{{{3^s}\sin \left( {\frac{\pi }{{2 \cdot {3^s}}}} \right)}}\notag\\
&\qquad\qquad = \frac{{\sin \left( {\frac{{{3^s}\pi }}{2}\frac{1}{{{3^s}}}} \right)}}{{{3^s}\sin \left( {\frac{\pi }{2}\frac{1}{{{3^s}}}} \right)}}\;\;\; = \left| {\frac{{\sin \left( {\frac{{{N_{{\rm{act}}}}\pi }}{2}\frac{1}{{{3^s}}}} \right)}}{{{N_{{\rm{act}}}}\sin \left( {\frac{\pi }{2}\frac{1}{{{3^s}}}} \right)}}} \right|,\notag\\
&\qquad \Rightarrow \qquad \left|{\sin \varphi _n - \sin \varphi _n^e} \right| = \frac{1}{{{3^s}}}\label{ce}.
\end{align}
By substituting (\ref{c3i}) into (\ref{phi}), the beam direction is given by
\begin{equation}
{\varphi _n} = \left\{ {\begin{split}
{\;\;\;\;\;\arcsin \left( {{3^{ - s}}(2n - 1) - 1} \right),\;\;\;{{\varphi _n}\in \rm{FR}}}\\
{\pi  - \arcsin \left( {{3^{ - s}}(2n - 1) - 1} \right),\;\;\;{\rm{otherwise}}}
\end{split}} \right..
\end{equation}
From (\ref{ce}), we obtain the coverage-edge direction as 
\begin{equation}
{\varphi _n^e} = \left\{ {\begin{split}
{\;\;\;\;\;\arcsin \left( {{3^{ - s}}(2n - 1\pm 1) \!-\! 1} \right),\;{{\varphi _n^e}\in \rm{FR}}}\\
{\pi  - \arcsin \left( {{3^{ - s}}(2n - 1\pm 1) \!-\! 1} \right),\;{\rm{otherwise}}}
\end{split}} \right..
\end{equation}
Thus, the beam coverage of ${\bm{\omega }}_n^s$ can be expressed as 
\begin{small}
\begin{equation}\label{cv2}
\begin{split}
&{\cal C}{\cal V}({\bm{\omega }}_n^s) = \\
&\left\{ {\begin{split}
{\left[ {\arcsin \left( {{3^{ - s}}(2n \!-\! 2) \!-\! 1} \right),\arcsin \left( {{3^{ - s}}  2n \!-\! 1} \right)} \right],\;\;{{\varphi _n^e}\in\rm{FR}}}\\
{\pi  \!-\! \left[ {\arcsin \left( {{3^{ - s}}(2n \!-\! 2) \!-\! 1} \right),\arcsin \left( {{3^{ - s}}  2n \!-\! 1} \right)} \right],{\rm{otherwise}}}
\end{split}} \right..
\end{split}
\end{equation}
\end{small}It can be similarly proved that (\ref{cv2}) still holds for $s=s_{\rm{max}}+1,...,\log _3^N$. As a result, it follows that
\begin{equation}
{\cal C}{\cal V}({\bm{\omega }}_n^s) = {\cal C}{\cal V}({\bm{\omega }}_{3n - 2}^{s + 1}) \cup {\cal C}{\cal V}({\bm{\omega }}_{3n - 1}^{s + 1}) \cup {\cal C}{\cal V}({\bm{\omega }}_{3n}^{s + 1}),
\end{equation}
which completes the proof.
\end{appendices}

\end{document}